\documentclass[12pt]{article}
\usepackage[utf8]{inputenc}
\usepackage[margin=15mm]{geometry}
\usepackage{amsmath,amsfonts,amsthm,amssymb}
\usepackage{float}
\usepackage{chngpage}
\usepackage{caption}
\usepackage{subcaption}
\usepackage{booktabs}
\usepackage{dcolumn}
\usepackage{breakcites}
\usepackage{siunitx}
\usepackage[title]{appendix}
\usepackage{graphicx}
\usepackage{url}
\usepackage{multirow}
\usepackage{siunitx}
\usepackage{pdflscape}
\usepackage{lineno}
\usepackage{todonotes}

\title{Transaction Cost Analytics for Corporate Bonds}

\author{
Xin Guo
\thanks{Department of Industrial Engineering and Operations Research, University of California, Berkeley, USA. Email: xinguo@berkeley.edu.}
\and Charles-Albert Lehalle
\thanks{Abu Dhabi Investment Authority (ADIA) and Imperial College London. Email: c.lehalle@imperial.ac.uk.}
\and Renyuan Xu
\thanks{Epstein Department of Industrial and Systems Engineering, University of Southern California, US \& Mathematical Institute, University of Oxford, UK. Email: renyuanx@usc.edu.}
}

\date{\today}

\providecommand{\keywords}[1]
{
  \small	
  \textbf{\textit{Keywords:}} #1
}
\begin{document}
\maketitle

\begin{abstract}
 Electronic platform has been increasingly popular for  executing large corporate bond orders by  asset managers, who in turn have  to assess the quality of their  executions via Transaction Cost Analysis (TCA). One of the challenges in TCA is to build a realistic benchmark for the expected transaction cost  and to characterize the price impact of each individual trade with given bond characteristics and market conditions. 

Taking the viewpoint of retail investors, this paper presents an analytical methodology for TCA of corporate bond trading.   Our  analysis is based on the TRACE Enhanced dataset; and starts with estimating the initiator of a bond transaction, followed by  estimating the bid-ask spread and the
mid-price dynamics. With these estimations, the first part of our study is to identify key features for corporate
bonds and to compute the expected average trading cost. This part is on the time scale of weekly transactions, and is by applying and comparing several regularized regression models.  The second part of our study is using the estimated mid-price dynamics to investigate the amplitude of its
price impact and the decay pattern of individual bond transaction. This part 
is on the time scale of each transaction of liquid corporate bonds, and is by applying a transient impact model  to estimate the price impact kernel using a non-parametric method.

Our benchmark model allows for identifying abnormal transactions and for enhancing  counter-party selections.  A key discovery of our study is the price impact asymmetry between customer-buy orders and consumer-sell orders.  
\end{abstract}

\keywords{Bond liquidity, transaction costs analysis, price impact, Enhanced TRACE, regression analysis, regularization method, data-driven decision making}

\newpage

\section{Introduction}
\label{sec:introduction}

Corporate bonds are critical to  firm finance and play an important part in asset management  \cite{bonds16nbis}.  Different from equity shares which are mostly traded via order books  available in multilateral trading facilities, cooperate  bonds are mainly traded via bilateral mechanisms  \cite{gueant16bond} due to  limited available electronic platforms \cite{citeulike:14555874}. 
 Even with  the introduction of the TRACE reporting system  in US in June 2002 and the establishment of MiFID 2 for electronic bond tradings in Europe in January 2018,  bond trading remains far less transparent than equity trading 
\cite{bess08bonds}.  

After the 2008 financial  crisis,  the macroprudencial regulation requires more transparency of corporate bond trading  to reduce information asymmetry  \cite{hender11bonds} between intermediaries and their clients, leading to an increase in capital requirement and in turn preventing banks  from taking large inventories as before \cite{citeulike:13795218}.  
This lower inventories, combined with the requirement of more transparency,  pushes banks and dealers towards flow driven business via electronification \cite{citeulike:13771761}.

In this new trading environment, asset managers, lacking pricing tools and private databases enjoyed by maker-dealers, have to assess the quality of corporate bonds execution via Transaction Cost Analysis (TCA). Details of TCA are then shared with the portfolio managers of investment firm to  review  market liquidity and for  allocation and hedging purposes \cite{re1}.

TCA is difficult  as there is a dire lack of benchmark \cite{collins1991methodology} for bond trading,  unlike equity trading where the bid-ask spread is an obvious and easy choice for the benchmark.
Instead, TCA needs to break down costs of  a particular bond trading according to brokers from all possible execution venues in fragmented markets, including order books, requests-for-quotes, voice, dark pools, and block discovery mechanisms.

\paragraph{Our work.} The goal of this  paper is to establish a TCA benchmark in bond trading for retail investors. That is, we take the standpoint of an individual investor to evaluate the execution performance of each transaction. 

Our TCA  is based on  the Enhanced TRACE  dataset from 2015 to 2016. We assume that TCA consists of  the bid-ask spread measuring the cost of illiquidity and the mid-price move measuring the impact of an individual trade.

Our analysis starts with a preliminary step of estimating the initiator of a bond transaction (Section \ref{sec:RPT}).
Initiator, currently missing from  the TRACE database,  indicates whether 
a given transaction is buyer-initiated or seller-initiated.
This estimated initiator of each trade enables us to  estimate the bid-ask spread and the mid-price dynamics (Section \ref{sec:RPT}).

With this preliminary step, the first part of our study is to  identify the most important features for corporate bonds and to compute  the expected average trading cost (Section \ref{sec:linreg}). This study is on the time scale of weekly transactions,  and is carried out via comparing several regularized regression models including the two-step Lasso, the Ridge regression, and the two-step Elastic Net regression. 
The response variable in the regression analysis is the estimated bid-ask spread from the preliminary analysis. 

Our regression approach manages to  select features for corporate bonds that are consistent with existing works, including the volatility of the bond price, the number of years from the issue date, and the \emph{activities} of the bond characterized by the number of trades and the traded amount (in dollars) per week. In addition, the number of trades and the traded amounts are found to play two opposite roles: the larger the amount traded in dollars, the smaller the bid-ask spread;  the more trades (for the same amount in dollars), the larger the bid-ask spread.  It is worth mentioning that  the R$^2$ value obtained from our regression analysis ranges from $0.50$ to $0.60$,  whereas the R$^2$ in existing works via regressions varies from $0.05$ to $0.20$ in \cite{hender11bonds}, $0.30$ to $0.50$ in \cite{larry07bond}, and $0.50$ to $0.80$ in \cite{dn12bond}.

The second part of our TCA is using the estimated mid-price dynamics to investigate the amplitude of its price impact and the price decay pattern of  individual bond transaction.
This study (detailed in Section \ref{sec:short}) is on the time scale of each transaction of liquid corporate bonds. It is done by applying a transient impact model (TIM) to estimate the price impact kernel via a non-parametric method. The transient impact functions  estimated in our study is found to share several important characteristics with those in the equity market:
\begin{itemize}
\item a price jump when the trade occurs,
\item a price decay after the initial jump,
\item  and the stabilization at a ``\emph{permanent level}''  higher than the initial price: this permanent impact can be interpreted as the informational content of the trade.
\end{itemize}
In addition, we discover an asymmetry in the amplitude of the initial price jump: \emph{buy-initiated transactions with more instantaneous impact than sell-initiated transactions} on corporate bonds. Note that such an asymmetry, not present in the equity market,  has  also been reported in  \cite{hender11bonds}  and \cite{ales16bond} for corporate bonds.

\paragraph{Existing works on TCA of corporate bonds.}
Empirical studies on transaction costs of corporate bonds are mostly from post-TRACE, as it was difficult for retail investors to obtain data in the pre-trace era. The post-TRACE trade reporting obligation started in US in July 2002. 
Because of the exogenous shock from the entry of TRACE, a number of earlier works \cite{gold07bond,ales16bond,bess08bonds}
 focused on the early years of its introduction in order  to identify the cost effect of this transparency.
Another family of post-TRACE studied the influence of adopting electronic and multilateral trading \cite{hender11bonds} and  decreasing borrowing costs from 2004 to 2007 \cite{asquith2013market}.

All these studies reached similar conclusions that the trading costs of corporate bonds decreased on average over the last twenty years. The  main  proxy for transaction costs adopted in these works was the (expected) bid-ask spread \cite{GlostenMilgrom85},  \cite{larry07bond}. Their main statistical approach was ordinary least square (OLS) regression to account for bond-specific or context-driven variations. The explanatory variables in these studies \cite{dn12bond,larry07bond,gold07bond,hel04bond} were the coupon, the maturity date, the number of years to maturity, the volatility, the risk-free rate, the expected recovery rate of the company, and the probability of default.\footnote{Note that \cite{citeulike:11289424} did not perform any linear regression, but relied on descriptive statistics, probably due to the lack of explanatory variables available during this period.} See Table \ref{tab:comp:literature} in Appendix \ref{sec:literature2} for a summary of the dataset and the years of bonds studied in these empirical analysis.

\paragraph{Existing works on asymmetric price impact.}
The price impact asymmetry between customer-initiated buy orders and customer-initiated sell orders in corporate bond market has been documented in the literature. For example,
this  asymmetry is reflected  in  the regression of Table IV of \cite{hender11bonds} since the coefficients of the buy and sell orders are  not of  the same amplitude for over-the-counter (OTC) trades (but not for electronic trades). 
 Such an asymmetry was also reported in Figure 15 of \cite{finra15bonds} and Table 1 of \cite{ales16bond}. The former
plotted the yearly average price change after five trades from 2003 to 2015, with the impact of buys  around 25\% more than the impact of sells.
The latter suggested that the average difference between the price of a transaction and the average price of the day is 56bp to 33bp for institutional buyers and -25bp to -21bp for institutional sellers on TRACE data from 2004 to 2012.

Our work is different from existing studies of average transaction costs with OLS, as we apply regularized regression models to select features within a broader class of candidate features. Unlike  previous studies on the price impact asymmetry from a static point of view,  we investigate the price impact curve of individual trade and analyze its  asymmetry in a dynamic setting,  characterizing both the amplitude of the price impact and the  decay via TIM models.
Moreover, the analysis and methodology presented in this paper are general and can be  applied to conduct TCA on other datasets including Standard TRACE.

\section{Data Processing and Bond Selection}
\label{data_pre}

\paragraph{Enhanced TRACE.}

TRACE, an acronym for {\it the Trade Reporting and Compliance Engine}, is the FINRA-developed mechanism that facilitates the mandatory reporting of over-the-counter secondary market transactions in eligible fixed income securities. 
TRACE database contains some useful (though limited) information and has been used for empirical studies by \cite{DICK2014} and \cite{citeulike:13771761}.

The main difficulty  working with TRACE is the lack of information on the liquidity offer. For example, there are neither quotes,  nor bid prices, nor ask prices. Instead, only final transactions are recorded, together with the type of the transaction: dealer-to-dealer, dealer-to-customer, or customer-to-customer.  
Therefore, besides TRACE, we also rely on Thomson Reuters to retrieve information  on the bonds traded, such as the amount issued, the coupon rate, the sector information,  rating information,
and Libor and Overnight Indexed Swap rate. In addition, we obtain the outstanding amount through the Mergent Fixed Income Securities Database (FISD).

There are two types of TRACE datasets, Standard TRACE dataset and Enhanced TRACE dataset.  Both TRACE  datasets contain corporate bond transactions. The difference is that transactions are available on Standard TRACE with a delay of two weeks, with  the volume of the transaction  capped at 1MM for high yield bonds and 5MM for  investment grade bonds. The Enhanced TRACE dataset has  uncapped volumes, with transactions  available with a delay of six months. There is   a separate dataset provided by FINRA for monthly price, return, coupon, and yield information for all corporate bonds traded since July 2002.

Our study is primarily based  on the Enhanced TRACE dataset.
We use the nontruncated transaction volumes on Enhanced TRACE along with other information from FISD and Thomson Reuters to construct the estimation of the bid-ask spread (Section \ref{preliminary}).  
It is worth noting that  the Enhanced TRACE dataset  and the Standard TRACE dataset  yield insignificant differences in terms of the estimation of the expected bid-ask spread, as shown in Section \ref{sec:two_trace_comparison}.

\paragraph{Data processing.}

The data used in our study is from January 1, 2015 to December 31, 2016, obtained from Wharton WRDS. During this period, there are 34,809,405 original trade reports, 390,193 reports of trade cancellations (approximately 1.1 percent of all original trade reports), 497,249 corrected trade reports (about 1.4 percent), and 28,005 reports of trade reversals. Trade reversals are transactions that have been changed after more than 20 days since they were initially recorded. Occasionally there are  multiple correction records for the same original trade
 and cancel records that  cancel  previously corrected trades.  There are 54,885 CUSIP\footnote{CUSIP stands for {\it Committee on Uniform Securities Identification Procedures}.}-days spread over 656 calendar days, many of which are weekends and holidays. The CUSIP-days are computed by {counting} all the trade days over all the CUSIP bonds.

In particular, for each transaction  of a bond, one can recover from Enhanced TRACE the following information:
 \begin{itemize}
\item $t^b_k$: the timestamp for the $k$th transaction of bond $b$;
\item $P^b_k$: the price of  the $k$th transaction of bond $b$;
\item $V^b_k$: the volume of the $k$th transaction of bond $b$;
\item the side of the dealer-to-customer transaction: customer buy order or customer sell order.
\end{itemize}
 
The data cleaning procedure combines the approaches of \cite{DICK2014} and \cite{citeulike:13771761}, as detailed in Appendix \ref{filtering}.
 In total, about 17.50\% reports are filtered out from the original Enhanced TRACE dataset. Among all the remaining 28,719,813 records, 14,071,375 (49\%)  are dealer-to-customer trades and the remaining 14,648,438 (51\%) are trades between dealers. 
These statistics are summarized in Table \ref{tbl:data_clean}.

After the data cleaning, appropriate bond selection is necessary to facilitate  the analysis of transaction costs,  for both the regression and  price impact analysis. 

\paragraph{Bond selection for regression analysis}\label{bond_selection_regression}

There are two types of bonds for regression analysis,  investment grade bonds and high yield bonds, which are picked from the standard universe of U.S. corporate bonds.  The investment grade bonds are selected from iShares iBoxx  Investment Grade Corporate Bond ETF, and the high yield bonds from the components of iShares iBoxx  High Yield Corporate Bond ETF. 
There are 1,033 current holdings of the former, among which {\color{black}538} bonds have more than one transaction recorded in Enhanced TRACE during the time period of Jan 1, 2015 to Dec 31, 2016.  There are 1,575 current holdings of the latter, 1485 of which have transaction records during the same period. Moreover, there are 
30 bonds that belong to both iShares iBoxx High Yield Corporate Bond ETF and iShares iBoxx  Investment Grade Corporate Bond ETF.
 The rating levels of all these 30 bonds  have been adjusted since issuance. Hence, there are a total of  1,993 bonds for the regression analysis.
 These selected bonds consist of 31.05 \% of the total 14,071,375 customer-to-dealer reports from  all bonds. 
Table \ref{tbl:data_clean}  reports this selection as ``Selection LR'' and Table~\ref{tbl:statistics_1993} reports the statistics of these selected bonds.

\begin{table}[H]
\centering\captionsetup{margin = 2.5cm}
\small{
\begin{center}
\setlength{\tabcolsep}{4pt}
\captionsetup{margin=2.5cm}
\hspace*{-4ex}\begin{tabular}{@{}l
*{4}{S[table-format=10.2,table-number-alignment=center]}@{}}
\toprule
{}&{Total}&{Investment Grade}&{High Yield}&{Bonds with rating changes}\\\midrule
    Number of trades& 4,371,363&3,102,791&1,109,177&159,395\\
    Number of customer buy &2,549,932&1,834,873&623,839&91220\\
    Number of customer sell &1,821,371&1,267,888&485,308&68175\\ 
    Total trading volume  (billion)&3,401.76 &2,438.26 &850.46&113.04\\
     Avg trading volume &778,202.926&785,834.74&766,775.08&709,163.43\\
     Avg price &102.05&104.26&97.81&88.58\\
    Prop volume of customer buy &55.3\%&56.8\%&51.7\%&58.2\%\\
    Prop volume of customer sell &44.7\%&43.2\%&48.3\%&41.8\%\\ \bottomrule
\end{tabular}
\end{center}}
\caption{\label{tbl:statistics_1993} Description of selected 1,993 bonds for regression (dealer-customer trades).}
\end{table}

 Note that our analysis throughout the paper focuses on trades between customers and dealers, which are  statistically different from  trades between dealers. The study for the latter requires the initiator analysis for dealer-to-dealer trades, which is infeasible to estimate given the current information from  the database.


\paragraph{Bond selection for price impact analysis.}\label{sec:price_imapct_bond_statistics}
  Given that the calculation of price impact curves requires a higher trading frequency,
 out of  all the 1,993 bonds for regression analysis,  the top-200 traded bonds (in terms of number of transactions) are selected to compute price impact curves. See the  ``Selection for PI''  step in Table \ref{tbl:data_clean} and the statistical summary of these 200 bonds in Table~\ref{tbl:statistics_1000}. Tables \ref{tbl:statistics_1993} and \ref{tbl:statistics_1000} show  that these top-200 bonds account for 32\% of the total number of transactions among the 1,993 bonds. Note that among the $30$ bonds with a rating level adjustment, $13$  belong to the top-200 traded bonds.

\begin{table}[H]
\centering\captionsetup{margin = 2.5cm}
\small{
\begin{center}
\setlength{\tabcolsep}{5pt}
\captionsetup{margin=2.5cm}
\hspace*{-4ex}\begin{tabular}{@{}l
*{4}{S[table-format=10.2,table-number-alignment=center]}@{}}
\toprule
{}&{Total}&{Investment Grade}&{High Yield}&{Bonds with rating changes}\\\midrule
    Number of trades&1,404,507& 980,005 &309,777&116,561\\
    Number of customer buy &836,825&588,128&180,378&68,319\\
    Number of customer sell &567,682&390,537&128,903&48,242	\\ 
    Total trading volume (billion)&605.5 &430.83 &107.70&66.98\\
     Avg trading volume &431,119.48&440,217.83&348,241.14&574,636.26\\
     Avg price &100.08&103.32&94.58&87.55\\
     Prop volume of customer buy &52.2\%&53.12\%&50.1\%&49.6\%\\
    Prop volume of customer sell &47.8\%&46.88\%&49.9\%&50.4\%\\ \bottomrule
\end{tabular}
\end{center}}
\caption{\label{tbl:statistics_1000} Description of the selected {\color{black}200} bonds for {\color{black}price impact analysis} (dealer-customer trades).}
\end{table}

\section{Preliminary Analysis}\label{preliminary}
Two key components for TCA   are the bid-ask spread and the mid-price dynamics, for which 
it is necessary to identify the  riskless-principle-trades (RPTs) and the initiator of a transaction. 

\subsection{RPT and Initiator}
\label{sec:RPT}
In the TRACE database, the information of the  initiator,  i.e., whether a given trade is a buyer-initiated  or a seller-initiated, is missing.  
In fact there is a substantial fraction of  transactions between dealers and customers  \cite{citeulike:13771761}  where the dealer has found two clients and put herself in between the transactions. These transactions   are called  \emph{riskless principal trades} (RPT) since the dealer  does not take any inventory risk by matching two clients. Consequently, it is not possible to recover the initiator of the RPT because there is no information on which of the two clients has initiated the trades.

Our first step is therefore to identify these RPTs. See Table~\ref{tbl:RPT-1993} and Table~\ref{tbl:nonRPT-1993} for the statistics of RPTs and non-RPTs.   Table~\ref{tbl:RPT} and Table~\ref{tbl:nonRPT} report the potential RPTs and non-RPT dealer-customer trades of the top-200 traded bonds. 
 See also Appendix \ref{sec:signtrades}  for a detailed literature review on RPT.\footnote{Our percentage of RPTs is lower than that reported in \cite{citeulike:13771761}, partly  because of different datasets with different time periods. \cite{citeulike:13771761} used the Standard TRACE dataset from April 1, 2014 to March 31, 2015, where the markers (``1MM+" and ``5MM+") for larger trades assign the same value to many large trades.  Finally, we only count the RPTs for a subset of bonds whereas \cite{citeulike:13771761} estimated the PRTs for a larger set of bonds.  }
 
After identifying and removing all the potential  RPTs, we consider the transaction initiated by the client. We define the \emph{sign of the transaction $\epsilon^b_k$} as $+1$ (i.e., ``buy'') if a client buys from a dealer and $\epsilon^b_k$ as
$-1$ (i.e., ``sell'') if a client sells to a dealer. When it is not possible to determine the sign of a trade as in the above RPT case, we assign $\epsilon_k^b$ to be zero.\footnote{Note that there are other  methods to estimate the sign of a trade when quote price is not available. See \cite{holthausen1987effect} for the tick test and \cite{lee1991inferring} for the inverse tick test.}

\begin{figure}[H]
  \begin{center}
    \includegraphics[width=0.6\columnwidth]{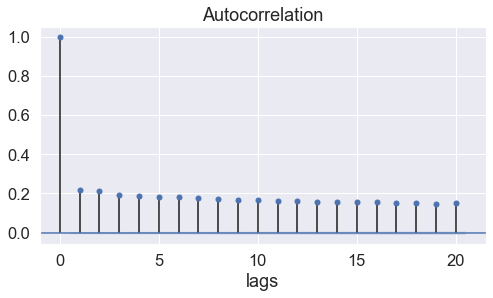}
    \caption{Auto-correlation of the signs.\label{fig:autocorrelation}}
  \end{center}
\end{figure}

 Figure \ref{fig:autocorrelation} reports the auto-correlation  of the order signs with lag 20. It suggests that with high confidence level, there is a persistent positive auto-correlation among order signs which delays very slowly.  In comparison, \cite{bouchaud2009markets} showed that the sign of market orders on equity market is strongly correlated in time.

\subsection{Bid-ask spread and mid-price estimation}

After identifying the initiator of each trade, we now analyze the two essential building blocks for TCA:  the bid-ask spread and the mid-price dynamics.

To start, let us find two consecutive trades that have opposite signs $\epsilon^b_{k+1}=-\epsilon^b_{k}$ with $\epsilon^b_{k}\ne 0$
and are sufficiently close in time (i.e., $|t^b_{k+1}-t^b_k|<\Delta t$). Let us then define the estimate of  bid-ask spread in absolute value as:
\begin{eqnarray} \label{estimated_vanilla_spread}
\psi^b_{k+1}:=(P^b_{k+1}-P^b_k)\cdot \epsilon^b_{k+1},
\end{eqnarray}
with $P_k^b$  the price of the $k$-th transaction of bond $b$.  We next estimate the mid-price at $t_k$ as:
\begin{equation}
M^b_k:=P^b_k - \epsilon^b_{k+1} \,\frac{\psi^b_{k+1}}{2}, 
\end{equation}
 and define the bid-ask spread in basis point (relative value) by
\begin{eqnarray} \label{estimated_spread}
s^b_{k+1}:=\frac{\psi^b_{k+1}}{M^b_k} \times 10000.
\end{eqnarray}
Note that the choice of   $\Delta t$ is 5-minute, which is largely due to the low
trading frequency of the corporate bond market.\footnote{$\Delta t<1$-minute is not realistic for the (illiquid) corporate bond market; meanwhile the choice of $\Delta t\ge 10$-minutes is infeasible due to lack of  data. Comparing results with $\Delta t=4$-minutes, $\Delta t=6$-minutes, and  $\Delta t=5$-minutes leads to the  choice of $\Delta t=5$-minute.} {Consequently, only 15.6\% of the transactions are used to calculate the bid-ask spread among bonds that are selected from Section \ref{bond_selection_regression}.} 

 We next check the reliability and stationarity of the estimated bid-ask spread. 
\paragraph{Reliability of the estimated spread.} 
We compare the estimated bid-ask spread with the one computed using bid and ask quotes provided by Composite Bloomberg Bond Trader (CBBT) for those bonds that are available in both the CBBT and Enhanced TRACE data sets. CBBT  is a composite price based on the most relevant executable quotations on FIT, Bloomberg's Fixed Income Trading platform. 
The CBBT pricing source provides average bid-ask prices based on executable quotes listed on Bloomberg's trading platform.
\cite{FGP2016} used the CBBT data as a measure of bond liquidity. {We only have access to} quote price data from Bloomberg CBBT 
from June 1, 2015 to  May 31, 2016 (12 months) for 2,361 investment grade bonds that belong to the iboxxIG universe, among which we have identified
1,401 bonds with records in both the Bloomberg CBBT database and the Enhanced TRACE subset.

Figure \ref{fig:spread_empirical} below shows the  plot for the empirical distribution of the spread from CBBT and the estimated spread from Enhanced TRACE for two arbitrarily chosen bonds, whose statistics are reported  in Table \ref{tbl:spread}.
It is noticeable (and expected)  that the CBBT spreads are larger than those estimated from real trades available in Enhanced TRACE. As \cite{FGP2016} pointed out, CBBT bid-ask spread estimates are based on quotes, and not on real transactions. As a consequence they include quotes that are not attractive enough (i.e., not small enough) to trigger a transaction. Since the bid-ask spread is the first component of implicit transaction costs, trades occur when they are smaller than the average bid-ask spread.

\begin{figure}[ht]
\centering
        \begin{subfigure}[b]{0.5\textwidth}
                \centering
                \includegraphics[width=\textwidth]{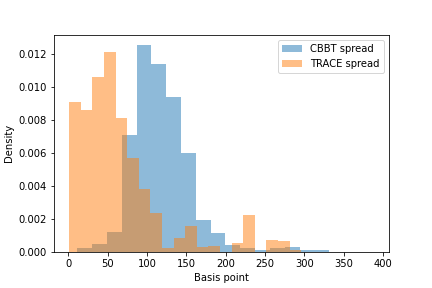}
                \caption{US375558BG78}
                \label{fig:gull}
        \end{subfigure}%
        \begin{subfigure}[b]{0.5\textwidth}
                \centering
                \includegraphics[width=\textwidth]{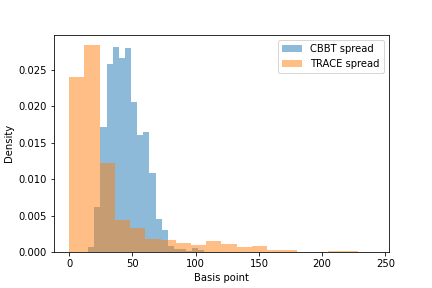}
               \caption{US126650CJ78}
                \label{fig:gull2}
        \end{subfigure}%
\caption{\label{fig:spread_empirical}Empirical distributions of the spread {\color{black}(Eqn \eqref{estimated_spread}}).}
\end{figure}

\begin{table}[ht]
\centering\captionsetup{margin = 2.5cm} 
\setlength{\tabcolsep}{8pt}
\captionsetup{margin=2.5cm}
\begin{tabular}{@{}l
*{6}{S[table-format=2.1,table-number-alignment=center]}@{}}
\toprule
{ } &{Coupon} & {Amount outstanding} & \multicolumn{2}{c}{Average spread (bp)} &
\multicolumn{2}{c}{Daily average} \\
  { } &{} & {} & \multicolumn{2}{c}{} &
\multicolumn{2}{c}{number of updates} \\ 
 \cmidrule(lr){4-5}\cmidrule(lr){6-7}         
{ } &{ }& {(USD)} & {CBBT} & {TRACE} &
{CBBT} & {TRACE}  \\ \midrule
US375558BG78 &4.6\% &\,\, 1,000,000,000 &118.78&66.24&2336.49&1.63\\
    US126650CJ78 &2.8\% &\,\,\,\,2,750,000,000&45.12&29.88&3406.20&11.50\\ \bottomrule
\end{tabular}
\caption{\label{tbl:spread}Spread comparison.}
\end{table}

\paragraph{Stationarity of the bid-ask spread.} We next check the consistency of the two approaches via a stationarity test on the ratio of the two estimates: the CBBT bid-ask spread and our trades-based estimates.  Denote $s^{\text{CBBT}}_{b,w}$ as the average spread for bond $b$ over the period $w$ taken from Bloomberg CBBT and $s^{\text{TRACE}}_{b,w}$ as the {average of the} estimated bid-ask spread for bond $b$ in period $w$ from Enhanced TRACE {\color{black}(see Eqn \eqref{estimated_spread})}. 
Similarly, define $R_{b,w}=s^{\text{CBBT}}_{b,w}/s^{\text{TRACE}}_{b,w}$ the ratio between these two spreads  for bond $b$ over the period $w$.

First of all, note that the empirical estimate of the bid-ask spread using Enhanced TRACE transactions are smaller than the CBBT ones: the average ratio is between 0.9 and 1 and its median is between 0.7 and 1, as summarized in Table \ref{tab:ratio}.

\begin{table}[H]
\centering\captionsetup{margin = 2.5cm} 
\setlength{\tabcolsep}{4pt}
\captionsetup{margin=2.5cm}
\begin{tabular}{@{}l
*{7}{S[table-format=6.2,table-number-alignment=center]}@{}}
\toprule
{Month }& {1 } & {2} & {3} & {4 } & {5 }&{6} \\ \midrule
Number of observations&1027&823&820&847&890&892\\
25\% & 0.7165&0.655&0.647&0.666&0.655&0.643\\
Median & 0.971 &0.931 & 0.899 & 0.900 & 0.852 &0.873\\
75\% & 1.238 & 1.213 & 1.224 & 1.182 &1.144 & 1.19\\
Mean & 0.970 & 0.956& 0.951& 0.941 &0.920& 0.931\\  \bottomrule
{Month}& {6} & {7} & {8} & {9}  &{10}&{12}\\ \midrule
Number of observations&939&914&1069&1062&1040&1047\\
25\% & 0.705&0.717&0.670&0.643&0.598&0.618\\
Median & 0.963&1.000&0.893&0.830&0.779&0.834\\
75\%& 1.289 & 1.309 & 1.182 & 1.090 & 1.056 & 1.130\\
Mean & 0.994 &1.019 &0.941 &0.889 &0.854 &0.893\\ \bottomrule
\end{tabular}
\caption{\label{tab:ratio}Statistics of the ratios.}
\end{table}

We also split the year 2016 into eleven groups of two consecutive months and check if this ratio is stationary from one period of the two months to the other. 
 We use two tests for the stationarity of  $R_{b,w}$.
The first is the one-way ANOVA test and the second is the  Kruskal-Wallis H-test.
{The former tests the stationarity of the mean and the latter tests the stationarity of the median.} The mathematical formulations and definitions of the ANOVA test and  the Kruskal-Wallis H-test are provided in Appendix \ref{a_h_tests}.

\begin{table}[H]
\centering\captionsetup{margin = 2.5cm} 
\setlength{\tabcolsep}{4pt}
\captionsetup{margin=2.5cm}
\begin{tabular}{@{}l
*{7}{S[table-format=6.2,table-number-alignment=center]}@{}}
\toprule
{Month }& { 1 and 2} & {2 and 3} & {3 and 4} & {4 and 5} & {5 and 6} & {6 and 7}\\ \midrule
ANOVA&4.59&0.10&0.39&1.71&0.52&15.32\\
(P-value)&0.032&0.751&0.533&0.191&0.472&0.000\\
H-test&3.564&0.163&0.154&1.890&0.296&15.16\\ 
(P-value)&0.038&0.686&0.6947&0.169&0.586&0.000\\ \bottomrule
{Month  }& {7 and 8} & {8 and 9} & {9 and 10}  &{10 and 11}&{11 and 12}\\ \midrule
ANOVA&2.21&25.54&14.15&6.72&7.81\\
(P-value)&0.136&0.000&0.0001&0.01&0.005\\
H-test    &2.30   &23.78   &12.03  &8.54&5.82\\ 
(P-value) &0.129  &0.000  &0.0005  &0.003&0.015\\ \bottomrule
\end{tabular}
\caption{\label{tab:compba}Results of ANOVA and Kruskal-Wallis H-tests.}
\end{table}

Table \ref{tab:compba} summarizes the results of both the ANOVA test and the Kruskal-Wallis H-test. With a $99\%$ confidence level, we accept (cannot reject) the null hypothesis (i.e., the  the ratios are stationary over time) 
 in both the ANOVA test and the Kruskal-Wallis test for $7$ of the total $11$ comparisons. We will thus use this estimated bid-ask spread in all  subsequent analyses because it can be operated over years of data using Enhanced TRACE, where CBBT is costly to obtain and linked to a private procedure owned by Bloomberg.  Nevertheless, these stationarity tests imply that a large investor using CBBT estimates could rely on the methodology presented thereafter and apply a ratio to interpret our results in terms of ``units'' of CBBT.

We have found similar results in terms of reliability and stationarity for the estimated mid-price dynamics, with details skipped here to avoid repetition.

\section{Regularized Regression Analysis for Bid-Ask Spread}
\label{sec:linreg}\label{sec:reglin}
\label{sec:linereg}

In this section, we use regularized regressions to identify the key features that drive the bid-ask spread, which provides the estimated cost for investors who needs to  move from one side (e.g., the buy side) to another side (e.g., the sell side). 

We will exploit several regularized regression models including OLS,  two-step Lasso, Ridge,  and two-step Elastic Net regressions, along with a $K$-fold cross-validation method,  to identify the most significant features and associated parameters for these models. 

As illustrated in Section \ref{data_pre}, there are total of $1,993$ bonds   selected for this regression analysis, along with a  total $152,408$ (weekly) samples processed from the Enhanced TRACE dataset during January $2015$ and December $2016$. 
Our regression analysis is performed on a weekly basis, with the weekly average bid-ask spread  computed according to Eqn. \eqref{estimated_spread}  serving as the response variable.

\subsection{Review of Methodologies}\label{sec:regression}
\noindent
We start by reviewing  the necessary notations  and steps for the regression analysis that will be used throughout the paper.

\paragraph{OLS.} 

OLS assumes that the regression function  is in linear form. That is,  given $\pmb{Y}:=(y_1,y_2,\cdots,y_n) \in \mathbb{R}^n$  the vector of $n$ observations of independent variables, and  $\pmb{X}:=(\pmb{1},\pmb{x}_1,\cdots,\pmb{x}_{w-1})$ with covariates $\pmb{1}\in \mathbb{R}^n$ and $\pmb{x}_i\in \mathbb{R}^n$ ($i=1,2,\cdots,w-1$), OLS is to find:
\begin{eqnarray}
\hat{\pmb{\theta}} := \arg \min_{\pmb{\theta} \in \mathbb{R}^w} \Big\{\|\pmb{Y}-\pmb{X}\theta\|^2_2\Big\}.
\end{eqnarray}
In an OLS,   $R^2$ is used to measure the goodness of fit for the model. 
 Meanwhile, an associated $p$-value  indicates the  significance level of the feature. 

\paragraph{Two-step Lasso.}
The first step   is to use Lasso regression to select the covariates by solving the following optimization problem:
\begin{eqnarray}\label{lasso}
 \min_{\pmb{\theta} \in \mathbb{R}^w} \left\{\frac{1}{N}\|\pmb{Y}-\pmb{X}\pmb{\theta}\|^2_2+ \lambda \sum_{j=1}^{w-1}|{\theta}_j|\right\}.
\end{eqnarray}
Here  a fixed constant  $\lambda$,  called the {\it  tunable hyperparameter}, controls both the size and the number of coefficients: a higher value of $\lambda$ leads to a smaller number of covariates in the linear model.
In the second step,  an OLS with only the selected covariates is applied
  \cite{BC2013}. That is, given the Lasso estimator $\pmb{\hat{\theta}}_l^{\lambda}$ in \eqref{lasso}, the subsequent OLS refitting is to find $\pmb{\bar{\theta}}^{\lambda}_l$ such that:
\begin{eqnarray}
\pmb{\bar{\theta}}^{\lambda}_l \in \arg\min_{\text{supp}[\pmb{\theta}]= \text{supp}[\pmb{\hat{\theta}}_l^{\lambda} ]} \Big\{\|\pmb{Y}-\pmb{X}\theta\|^2\Big\}.
\end{eqnarray}
 $\pmb{\bar{\theta}}^{\lambda}_l$ is thus called the estimator for the LSLasso (least-squares Lasso), also known as post-Lasso.
This two-step Lasso estimation procedure has been shown to produce a smaller bias than Lasso for a range of models \cite{BC2013}, \cite{LEDERER2013}, and \cite{CLS2017}.

\paragraph{Ridge regression.} The penalty term in the Ridge regression is 
of the $L_2$ norm. That is, for a fixed hyperparameter $\lambda$, Ridge regression  is to solve for:
\begin{eqnarray}\label{ridge}
\pmb{\hat{\theta}}^{\lambda}_r \in \arg \min_{\pmb{\theta} \in \mathbb{R}^w} \left\{\frac{1}{N}\|\pmb{Y}-\pmb{X}\pmb{\theta}\|^2_2+ \lambda \sum_{j=1}^{w-1}{\theta}_j^2\right\}.
\end{eqnarray}

\paragraph{Two-step Elastic Net regression.}
Elastic Net (EN) regression, introduced in \cite{citeulike:10520959}, is a hybrid of Lasso and Ridge. That is, for a fixed hyperparameter
 $(\lambda,\alpha)$ with $\alpha \in [0,1]$, EN is to solve for:
\begin{eqnarray}
\pmb{\hat{\theta}}^{\lambda}_e \in \arg \min_{\pmb{\theta} \in \mathbb{R}^w} \left\{\frac{1}{N}\|\pmb{Y}-\pmb{X}\pmb{\theta}\|^2_2+ \alpha\lambda\sum_{j=1}^{w-1}|{\theta}_j|+(1-\alpha)\lambda \sum_{j=1}^{w-1}{\theta}_j^2\right\}. \label{eq:en}
\end{eqnarray}
Note that the Lasso regression is recovered from Eqn. \eqref{eq:en} by taking $\alpha = 1$ and the Ridge regression is recovered by taking $\alpha=0$. Similar to the two-step Lasso, the second step of the Two-step Elastic Net  is to fit an OLS with only the selected covariates. That is, given the EN estimator $\pmb{\hat{\theta}}^{\lambda}_e$ in \eqref{eq:en}, the subsequent OLS refitting is to find $\pmb{\bar{\theta}}^{\lambda}_e$ such that:
\begin{eqnarray}
\pmb{\bar{\theta}}^{\lambda}_e \in \arg\min_{\text{supp}[\pmb{\theta}]= \text{supp}[\pmb{\hat{\theta}}_e^{\lambda} ]} \Big\{\|\pmb{Y}-\pmb{X}\theta\|^2\Big\}.
\end{eqnarray}

\paragraph{Cross-validation.}
 In all three regularized regression models,  the selection of  hyperparameters  is by the standard K-fold cross-validation approach to improve the predictive power of the model. That is,  the dataset is randomly divided into $K$ subsets. Each time, one of the $K$ subsets is used as the validation set and the remaining $K-1$ subsets form a training set. In this approach, every data point is in a validation set exactly once and in a training set $K-1$ times. The variance of the resulting estimate is reduced as $K$ increases.

\paragraph{Out-of-sample test.}
With the hyperparameters selected from the cross-validation step,  the coefficients for a regression model are estimated using  the training and validation datasets.  The performance of the regression model is then evaluated with the test dataset. For financial applications, the time period of the test dataset needs to be after those of the training and validation datasets to ensure the information adaptiveness.

Given a test dataset $(\widetilde{\pmb{X}},\widetilde{\pmb{Y}})$ with size $m$ and $\widetilde{\pmb{Y}} = (\tilde{y}_1,\cdots,\tilde{y}_m)$,
 the following relative error function is used as the criterion to measure the performance:
\begin{eqnarray}\label{eq:relative_error}
{\rm Relative \,\,err} = \frac{1}{m}\sum_{j=1}^{m}\frac{|\tilde{y}_j-\hat{y}_j|}{|\tilde{y}_j|},
\end{eqnarray}
where $\tilde{y}_{j}$ is the true label and $\hat{y}_j$ is the label predicted from the regression model for test sample $j$ ($j=1,2,\cdots,m$).

\subsection{Features for Regression Analysis}
The features in the regression consist of  two categories. One category concerns bond information, including time to maturity date, time since issued date, coupon rate, amount outstanding, and duration. The other category focuses on trade information including average transaction price, volatility, proportion of costumer-buys (sells), LIBOR-OIS rate, and the 5-year treasury rate during the given week. More specifically, we consider:
\begin{itemize}
\item {\bf Volatility}: calculated from the trade price. For bond $b$, assume there are $n$ trades in week $w$. Recall $P^b_j$ as the trade price of the $j^{th}$ transaction ($j=0,1,2,\cdots,n$) of bond $b$. Denote the log return $r_{i}^b = \log( \frac{P^b_{i}}{P^b_{{i-1}}})$ ($i=1,2,\ldots,n$) and the average return $\bar{r}^b = {\sum_{i=1}^n r_i^b}/{n}$. Then the  volatility in week $w$ is:
{$$\sigma_b = \sqrt{  \frac{1}{n-1} \sum_{i=1}^n ( r_i^b-\bar{r}^b)^2  }\cdot 100.$$}
Notice that $n$ may vary from bond to bond and from week to week.
\item {\bf Number of trading days}: the number of days that bond $b$ is traded {during the week}.
\item {\bf Log(zero trade days)}: the log of the number of days that bond $b$ is  not traded {during the week}.
\item {\bf Proportion of buy/sell number}: estimated by counting the number of customer-buy orders and the number of customer-sell orders in week $w$ and calculating the proportion of buys and sells for each bond $b$.
\item {\bf Proportion of buy/sell volume}: estimated by taking the total volume (in dollars) for customer-buy orders and customer-sell orders in week $w$ and calculating  the proportion of buys and sells for each bond $b$.
\item  {\bf Trading activity}:  the log of the {\it number of trades} in the week.
\item {\bf Total volume}:  the weekly total trading volume {in dollars} of both customer-dealer trades and dealer-dealer trades.
\item {\bf Average price}: the weekly average trade price {in dollars}.
\item {\bf Coupon}: annual coupon payments paid by the issuer relative to the bond's face or par value. The coupon rate is the yield the bond paid on its issue date. This yield changes as the value of the bond changes, thus giving the bond's yield to maturity.
\item {\bf Duration}: an approximation of a bond's price sensitivity to changes in interest rates which is defined as:
\begin{eqnarray*}
D^b = \sum_t \frac{\mbox{PV}(C^b_t)}{ \sum_t \mbox{PV}(C_t^b)} \times t
\end{eqnarray*}
for bond $b$, where $C_t^b$ is the cash flow on date t, $PV(C^b_t)$ is its present value (evaluated at the bond's yield), and $\sum_t \mbox{PV}(C_t^b)$ is the total present value of the cash flow, which is equal to the bond's current price.
\item {\bf Years to maturity}:  the time to maturity date calculated in years.
\item {\bf Years since issuance}: the time since issued  date counted in years.
\item {\bf Amount outstanding}: the principal amount outstanding of a bond;  sometimes referred to as the notional amount.
\item {\bf Turnover}: { the volume of bonds {\color{black}traded} relative to the total volume of outstanding bonds.} The inverse of the turnover can be interpreted as the average holding time of the bond. For instance, a  turnover of one implies an average holding time of about two weeks.
\item {\bf LIBOR-OIS rate}:  London inter-bank offer rate (LIBOR) is  the rate at which banks indicate they are willing to lend to other banks for a specified term of the loan: Overnight indexed swap (OIS) rate is the rate on a derivative contract on the overnight rate. The term LIBOR-OIS spread is assumed to be a measure of the health of banks because it reflects the  default risk associated with lending to other banks.  In this analysis, the 1-month LIBOR-OIS rate is used to indicate the bank health condition over time.
\item {\bf Indicator of high yield (HY) or investment grade (IG) bond}: indicator of  the bond .

\item {\bf Indicator of different sectors}: including nine different sectors such as basic materials sector (S1), communications sector (S2), consumer \& cyclical sector (S3), consumer \& non-cyclical sector (S4), energy sector (S5), financial sector (S6), industrial sector (S7), technology sector (S8), and utilities sector (S9).

\end{itemize}
Table \ref{tab:features_stats} provides descriptive statistics of these response variables and features.

\begin{table}[H]
  \centering\captionsetup{margin = -3.5cm}
  \setlength{\tabcolsep}{0pt}
  \begin{tabular}{lrrrrr}
    & \rule{2ex}{0pt}Mean & \rule{4ex}{0pt}std. & \rule{3ex}{0pt}q-25\% & \rule{2ex}{0pt}Median & \rule{2ex}{0pt}q-75\%\\\hline
    Bid-ask spread (bp) & 68.79 & 69.1 & 23.14 & 45.82 & 90.91 \\[0.2em]
    Volatility  &  0.785 & 0.582 & 0.388 & 0.646 & 1.039\\
    Trading activity& 1.28 & 0/35 & 1.04 & 1.26 & 1.51\\
    Log(\$ traded volume) & 6.90 & 0.63 & 6.50 & 6.94 & 7.33\\
    Nbe trading days & 4.41 & 0.80 & 4.00& 5.00& 5.00\\
    Log(zero trade days) & 0.12 & 0.18 & 0.00& 0.00 &0.30\\
    Prop. nbe buy & 0.45 & 0.18 & 0.33 & 0.45 & 0.57\\
    Prop. nbe sell & 0.55 & 0.18 & 0.43 & 0.55 & 0.67 \\
    Prop. \$ buy & 0.49 & 0.25 & 0.32 & 0.50 & 0.66\\
    Prop. \$ sell & 0.51 & 0.25 & 0.34 & 0.51 & 0.69\\
    Avg. price & 103.69 & 10.93 & 100.06 & 102.87 & 107.93\\
    Coupon & 4.69& 1.78 & 3.38 & 4.65 & 5.93\\
    Duration & 5.72 & 4.21& 2.75& 6.0& 8.0\\
    Years to maturity & 8.23 & 7.91& 3.0 & 6.0 & 8.0 \\
    Years since issuance & 3.99 & 2.93 & 2.0&3.0&5.0\\
    Turnover ($\times 10^{-2}$)& 1.80 & 6.10 & 0.30 & 0.80 & 1.90\\
    LIBOR-OIS & 0.21 & 0.09 & 0.14 & 0.20 & 0.25 \\
  \end{tabular}\\[.5em]
  \begin{tabular}{lrrclrr}
    & \rule{2ex}{0pt}Mean & \rule{4ex}{0pt}std. &\rule{4ex}{0pt}&& \rule{2ex}{0pt}Mean & \rule{4ex}{0pt}std. \\\cline{1-3}\cline{5-7}
    High yield & 0.27 & 0.44 & & Invest. grade & 0.73 & 0.44\\[0.2em]
    S: Basic Material & 0.03 & 0.18 && S: Communications & 0.15 & 0.38\\
    S: Consumer, Cyclical & 0.11 & 0.31 && S: Consumer, Non-cyclical & 0.14 & 0.35 \\
    S: Energy & 0.16 & 0.37 & & S: Financial & 0.27 & 0.44 \\
    S: Industrial & 0.07 & 0.25 && S: Technology & 0.07 & 0.25\\
    S: Utilities & 0.01 & 0.11 && & &\\
  \end{tabular}
   \caption{\label{tab:features_stats}Statistics of the response variable and the features.}
\end{table}


\paragraph{Hyperparameter selection.}
Specific to the regression models aforementioned, denote $\pmb{\mu}$ as the parameter for one of the regression models (for example, $\pmb{\mu} =(\lambda_e,\alpha)$ for EN).  Partition in log-scale is used for $m$ different hyperparameter values for $\pmb{\mu}$ and  the training dataset is divided  into $K$ folds for   cross-validation. For each {leave-out} fold $i$,    $R^2_i({\pmb{\mu}})$ is computed with regression coefficients calculated using the other $K-1$ folds. Hence for each $\lambda$, there is  an empirical distribution of  $\tilde{R^2}({\pmb{\mu}}) = \{R^2_i({\pmb{\mu}}), i=1,2,\cdots,K\}.$ Denote ${\widehat{R^2}}(\pmb{\mu})$ and $\sigma_{R^2}(\pmb{\mu})$ as the mean and standard deviation of the empirical distribution with parameter $\pmb{\mu}$, and define the confidence interval by:
\begin{eqnarray}\label{ci_1}
  {\frak{I}}_1(\pmb{\mu}) =\left[ {\widehat{R^2}}(\pmb{\mu}) - {\sigma_{R^2}(\pmb{\mu})\over\sqrt{K}}, %
  {\widehat{R^2}}(\pmb{\mu}) + {\sigma_{R^2}(\pmb{\mu})\over\sqrt{K}}\right].
\end{eqnarray}
Then  $\pmb{\mu}$ is picked such that the number of $ \tilde{R^2}(\pmb{\mu})$ in $ {\frak{I}}_1(\pmb{\mu})$ is maximized. Moreover, define:
\begin{eqnarray}\label{ci_2}
{\frak{I}}_2(\pmb{\mu}) = \left[ {\widehat{R^2}}(\pmb{\mu}) - {\sigma_{R^2}(\pmb{\mu})}, %
  {\widehat{R^2}}(\pmb{\mu}) + {\sigma_{R^2}(\pmb{\mu})}\right].
\end{eqnarray}
Note that ${\frak{I}}_2(\pmb{\mu})$ in \eqref{ci_2} is a relaxation of ${\frak{I}}_1(\pmb{\mu})$ in \eqref{ci_1}. When the number of $ \{R^2_i({\pmb{\mu}})\}$ is not sensitive to $\pmb{\mu}$ in ${\frak{I}}_1(\pmb{\mu})$, one can compare ${\frak{I}}_2(\pmb{\mu})$ instead.

The training dataset consists of data from  January 1, 2015 to December 31, 2016, and the test dataset (for out-of-sample performance) consists of data from  January 1, 2017 to March 31, 2017.

\subsection{Features Identified by Regression Analysis }
In this section, we present the results from the  two-step Lasso, the Ridge,  and the EN regressions, including  the most significant features and associated parameters identified by these models.  

\subsubsection{Benchmark model: OLS}
 We first  summarize the result from the benchmark method OLS. As seen  in Table \ref{OLS}, all but two of the estimated coefficients are statistically significant at any reasonable level of significance. 
  The two exceptions are {{\it year to maturity} and {\it turnover}}. Moreover,
\begin{itemize}
{
\item {The coefficients of {\it Prop number of buys} and {\it Prop number of sells} have the same sign but different values. The coefficient of {\it Prop number of buys} is roughly one third of the coefficient of {\it Prop number of sells}.
Similarly, both of the coefficients of {\it Prop buy volume} and {\it Prop sell volume} are positive. The coefficient of {\it Prop buy volume} is roughly half of the coefficient of {\it Prop sell volume}.}
This shows the asymmetry between customer buy orders and customer sell orders. This is consistent with  numerous studies, i.g., \cite{FGP2016}, suggesting that dealers offer tighter quotes for larger trades than for smaller ones.  
\item {\it Avg price} has a small effect on the bid-ask spread.
\item The indicators of different sectors have different coefficients, but the overall values are small.
\item The {\it Log(Total volume)} coefficient is negative as expected. With value $-21.4028$, the estimated coefficient implies that a  increase of 10,000 in trade size  would make a retail-size trade into a large institutional-size trade and would reduce the bid-ask spread by {100 basis points}.
\item The {\it Indicator of investment grade bonds} coefficient is negative and the \emph{Indicator of high yield bonds} coefficient is positive. This is consistent with the well-documented empirical findings: larger spreads for high yield bonds and smaller spreads for investment grade bonds.}
\end{itemize}

\subsubsection{Features Identified from Two-Step Lasso}
We then present the results from a class of two-step Lasso parameterized by different $\pmb{\mu}=\lambda_{l}$ and discuss how to select the best $\lambda_{l}$ with cross-validation.

In this analysis,  20 different values of $\pmb{\mu}=\lambda_{l}$ are picked with a partition in the range of $[10^{-1},10^{3}]$ in the log scale. 
Note that the ranges of hyperparameters are different for two-step Lasso, Ridge, and two-step EN, shown in the figures of cross-validation scores  (Figures ~\ref{fig:cross_validation}, ~\ref{fig:cross_validation_l2}, and ~\ref{fig:cross_validation_EN}).  The range is selected according to the sensitivity of  the model  with a larger prior partition grid.

Figure~\ref{fig:cross_validation} shows the 25\%, 50\%, and 75\% percentiles of out-of-sample $\tilde{R}^2$ with different $\lambda_l$ values. One can see that all three 25\%, 50\% and 75\% curves decrease fast before {$\lambda_l^*=2.98$} and tend to be flat after $\lambda_l^*$. Also, both $\frak{I}_1(\lambda_l)$ and $\frak{I}_2(\lambda_l)$ are large when $\lambda_l=  \lambda_l^*$. Hence, $\lambda_l^*$ is a good choice of the regularization level.  Table \ref{tab:R2_count_lasso} shows
 $\lambda_l$'s along with $\frak{I}_1(\lambda_l)$ and $\frak{I}_2(\lambda_l)$, in which the number of $\tilde{R}^2$ are the largest, respectively.  Table \ref{LASSO_normal_scale} shows the features selected from the first step of the two-step
 Lasso, with corresponding parameters {$\lambda^*_l = 1.13,$ $2.98$, and $7.85$}, respectively.
 It also shows the models from the OLS regression in the second step of
 the two-step Lasso. For instance,
 in Model L2 of Table \ref{LASSO_normal_scale} with $\lambda^*_l = 2.98$, the model is of the form with four features such that:
\begin{eqnarray}\label{eq:two_step_lasso_model}
\text{Bid-ask spread} =&&83.48\times \text{ Volatility}+39.07 \times \text{ Trading activity}\nonumber\\
&-&19.45\times  \text{Log(Total volume)} +0.23\times \text{ Issued years}+86.
\end{eqnarray}

In addition, as seen  from Table~\ref{LASSO_normal_scale}:

\begin{itemize}
\item The coefficient of {\it Volatility } is positive with value $83.48$. This is consistent with existing  theoretical and empirical studies in 
 market microstructure in  that a higher return volatility is predicted to lead to decreased liquidity (e.g., \cite{Stoll1978}).  
\item The coefficient of {\it Issued years} is positive with value {$0.23$}, which means a newly issued bond will have a small bid-ask spread. This is consistent with the work of \cite{KNP2016}, which argued that recent and large issues are cheaper to trade than seasoned and small ones.
\item The number of trades per day $N$ and the trade volume $V$ (in dollars)  suggest a joint impact of order {\color{black}$\log({N}/{\sqrt{V}})$} on the bid-ask spread.
Section \ref{regression_conclusion} provides  a detailed analysis of this relationship.
\item Finally, $\lambda_l=7.85$ leads to the features of {\it Volatility} and  {\it Issued year} in model L3. Compared to model L2 with four features and $R^2=51.5\%$,  $R^2$ in EN3 drops to $43.22\%$. In our view,  the model with four features, significantly  reduced from the original $26$ features, is preferable to EN3.
\end{itemize}

It is worth noting that our findings are  supported by previous studies. For instance, \cite{CMMS2005} found that credit quality, the age of a bond, the size of a bond issue, the original
maturity value of a bond at issuance date, and provisions such as a call, put, or convertible options all have strong impacts on liquidity. {\color{black}\cite{choi2019customer} showed that trade size and  maturity date are important features to understand in the bid-ask spread. }

\subsubsection{Features Identified from Ridge}

We next discuss the results from a group of Ridge regression methods parameterized by different $\lambda_r$ values and analyze how to select the best $\lambda_r$ with cross-validation.

In this analysis, 20 different values of $\lambda_r$ are chosen in the range of $[10^2,10^{8}]$ with uniform partition in the log scale. 
Figure~\ref{fig:cross_validation_l2} shows the 25\%, 50\% and 75\% percentiles of out-of-sample $\tilde{R}^2$ with different $\lambda_r$ values. One can see that all three 25\%, 50\%, and 75\% curves start to decrease at {$\lambda_r^*=1.27\cdot 10^{6}$}. Hence {$1.27\cdot 10^{6}$} is a good choice for the regularization level.

 Table \ref{tab:R2_count_ridge} shows
$CI(\lambda_r)$ and $CI_2(\lambda_r)$ for different values of  $\lambda_r$, in which the number of $\tilde{R}^2$ are the largest, respectively. Table \ref{Ridge_normal_scale} shows the results of Ridge regressions with parameters  {$\lambda^*_r = 1.62\cdot 10^{4}, 6.95\cdot 10^{4}$, and $1.27\cdot 10^{6}$}.

The analysis by the Ridge regression is consistent with the findings from the two-step Lasso. In particular,
\begin{itemize}
\item When $\lambda_r$ goes up, the coefficients of the following features go to $0$ very fast: {\it Indicator functions of different sectors}, {\it Proportion of buy (or sell) volumes (or numbers)} , {\it Turnover }, and {\it Number of trading days}. Note that from Table \ref{Ridge_normal_scale} these features are also excluded from Model L3 and L4 of Table~\ref{LASSO_normal_scale}, which means that results from these two approaches are consistent.
\item  When $\lambda_r$ takes a large value {$6.95\cdot 10^{4}$}, the {\it Volatility},  {\it Issued years},  {\it Trading activity} and {\it Log(Total volume)} are still significant. This is also consistent with the findings from Lasso in Table~\ref{LASSO_normal_scale}. 
\item Both two-step Lasso and Ridge regressions point to the significance of the time value and  the special structure of bonds. The variable {\it years since issuance} is significant in two of the two-step Lasso models, L1 and L3, and all three Ridge regression models. 
\end{itemize}
The difference between Lasso and Ridge regression:
{\it Avg price} is not significant in all three two-step Lasso models, whereas it is significant in all three Ridge models. This inconsistency is expected because of the collinearity among features.
When features are correlated, Lasso tends to select one feature from a group of correlated features while Ridge  tends to penalize the group of correlated features towards the same coefficients \cite{citeulike:10520959}. Indeed, as shown in Models R1, R2 and R3, the coefficients of {\it Prop number of buys} and {\it Prop number of sells} have the same value but different signs; the coefficients of {\it Prop buy volume} and {\it Prop sell volume} also have the same value but  different signs.
Additionally, the reappearance of {\it Avg price} in Model EN3 is due to this group effect as well.

\subsubsection{Features Identified from Two-Step EN}
Finally, we  discuss the results from the two-step EN models parameterized by different $\lambda_e$ values and analyze how to select the best $\lambda_e$ with cross-validation. 

Figure~\ref{fig:cross_validation_EN} shows the 25\%, 50\%, and 75\% percentiles of out-of-sample $\tilde{R}^2$ with different $\lambda_e$ values given different $\alpha=0.2,0.5$ and $0.8$. 
When {$(\alpha,\lambda_e)=(0.5,10^{3})$ and $(\alpha,\lambda_e)=(0.8,10^{3})$}, more than 170 empirical $\tilde{R}^2$ falls into $\frak{I}_2$. This is because the hyperparameter over-penalizes the model  such that all the coefficients and the empirical $\tilde{R}^2$ are all nearly zero. Therefore, these sets of hyperparameters should be excluded.

Instead, for the analysis the following parameters are selected $(\alpha,\lambda_e)=(0.5,0.774)$, $(\alpha,\lambda_e)=(0.8,2.15)$, and $(\alpha,\lambda_e)=(0.5,129)$.
Parameter  $(\alpha,\lambda_e)=(0.8,2.15)$ leads to the following set of features: {\it Volatility}, {\it Number of trades}, {\it Log(Total volume)} and {\it Issued year}. This is consistent with the feature selection in the two-step Lasso model L2.
$(\alpha,\lambda_e)=(0.5,129)$ leads to the features: {\it Volatility} and  {\it Average price} in model EN3. Compared with model EN2 with four features and $R^2=51.5\%$, the $R^2$ in EN3 drops to $42.62\%$. Similar to the argument for $L3$, model EN2 with four features is superior to  EN3, in our view.
 
Through all three different regression models, the consensus is that {\it Volatility} and {\it Issued years} are important features.

\subsubsection{Out-of-sample Performance}\label{sec:oos}

It is well recognized   out-of-sample forecast performance is generally considered more trustworthy than  in-sample performance \cite{alpaydin2020introduction}.
 The latter can be more sensitive to outliers and data mining, while the former tends to better reflect the information available to the forecaster in ``real time''.
 
\begin{figure}[H]
\centering
        \begin{subfigure}[b]{0.45\textwidth}
                \centering
                \includegraphics[width=\textwidth]{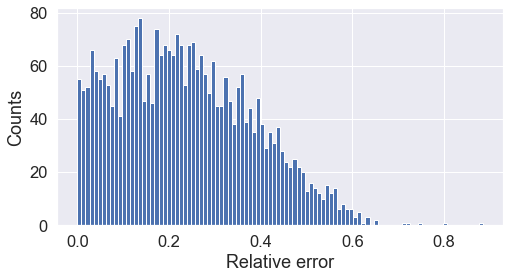}
                \caption{OLS model.}
                \label{fig:OOS_OLS}
        \end{subfigure}
         \begin{subfigure}[b]{0.45\textwidth}
                \centering
                \includegraphics[width=\textwidth]{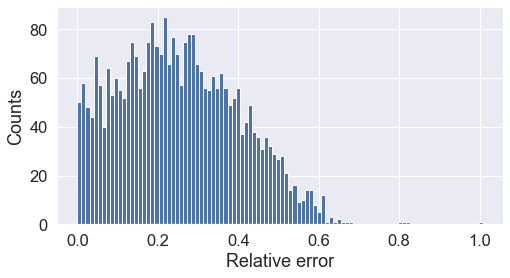}
                \caption{Two-step LASSO Model L2 with parameter $\lambda_l = 2.98$.}
                \label{fig:OOS}
        \end{subfigure}
        
         \begin{subfigure}[b]{0.45\textwidth}
                \centering
                \includegraphics[width=\textwidth]{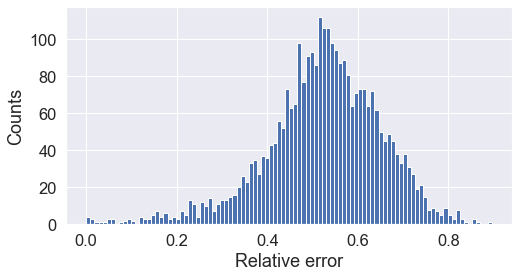}
                \caption{Ridge Model R3 with hyper-parameter \\$\lambda_r = 1.26\times 10^{6}.$}
                \label{fig:Ridge}
        \end{subfigure}%
         \begin{subfigure}[b]{0.45\textwidth}
                \centering
                \includegraphics[width=\textwidth]{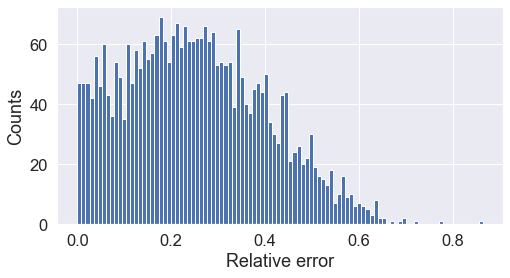}
                \caption{Two-step EN Model EN1 with hyper-parameter $(\alpha,\lambda_e) = (0.5,0.77)$}
                \label{fig:EN}
        \end{subfigure}%
\caption{\label{fig:relative_error} Out-of-sample test on data during  January-March 2017.}
\end{figure}

 In this subsection, we test the out-of-sample performance of all  regression models  on an unseen dataset during the period of January-March 2017.
 The distributions of the relative errors (defined in \eqref{eq:relative_error}) are provided in Figure \ref{fig:relative_error}  and the mean of the relative errors are recorded in Table \ref{tab:OOS_erros}.
For the OLS model, errors  smaller than $0.3$ account for more than $70\%$ of  the testing dataset. Similar results hold for the two-step LASSO and two-step EN for which  errors  smaller than $0.3$ account for more than $60\%$ of the testing dataset. The errors are bigger for Ridge, as expected, since the coefficients are biased.

The value of $R^2$ in Table \ref{OLS} indicates that the OLS model can explain around 60\% of the variance hence its out-of-sample performance is acceptable. Similar results are obtained from the two-step LASSO model EN1 and the two-step LASSO model L2. Ridge regression model R3 has a larger mean relative error since the coefficients are biased with the L2 penalty term, which is also expected.

\begin{table}[H]
\centering\captionsetup{margin = 2.5cm} 
\setlength{\tabcolsep}{4pt}
\captionsetup{margin=2.5cm}
\begin{tabular}{@{}l
*{5}{S[table-format=6.2,table-number-alignment=center]}@{}}
\toprule
{Model }& { OLS } & {L2} & {R3} & {EN1} \\ \midrule
Mean relative error &0.222&0.249&0.522&0.252\\ \bottomrule
\end{tabular}
\caption{Mean relative error of out-of-sample test on data  during  January-March 2017.\label{tab:OOS_erros}}
\end{table}

\subsection{Summary:  Bid-Ask Spread of Corporate Bonds}
\label{regression_conclusion}

Different linear regressions are performed for feature selection and for estimating the  bid-ask spread using two kinds of variables: one describing the bond and the other characterizing the market.
   Tables \ref{OLS}-\ref{Enet_normal_scale} summarize the results, with comparisons to the benchmark OLS approach. 

These regressions allow for computing an ``expected bid-ask spread'' for a given week, which can be used as a benchmark cost for TCA. In particular, 
  \begin{itemize}
  \item \emph{Volatility} is an important feature, as expected by both empirical observations and theory:
    the larger the volatility, the larger the bid-ask spread. It has been  observed in practice that an increase of 5\% in volatility, that is $1/2$ of its standard deviation in our dataset, corresponds to an increase of the bid-ask spread by 25 basis points, which is around one third of its standard deviation.
  \item \emph{Number of trades per day} $N$ and \emph{Traded volume} $V$ (in dollars) are both important variables (in log units), with coefficients suggesting  {\color{black}${\rm log}(N/\sqrt{V})$} being the feature  impacting the bid-ask spread in the basis points,\footnote{
  The ridge regression suggests  the feature being in the form of  $\frac{N}{V}$, with an addition term of the average price. Possible explanation: penalization in the Ridge regression  tends to avoid large coefficients in the regression.} implying that:
    \begin{itemize}
    \item for a given \emph{trading activity} $N$, the larger the traded volume, the smaller the bid-ask spread (in basis points);
    \item for a given traded volume in dollars, the lower the average trade size (i.e., the more trades), the larger the bid-ask spread. 
    \end{itemize}
Our result is compatible with the documented stylized fact that for corporate bonds: small trade size obtain a worse bid-ask spread than large trades  \cite{FGP2016}.
  \item The value of the coupon and the duration of the corporate bond play a small role in the formation of the bid-ask spread (both with a positive coefficient).
  \item Last but not least, the \emph{Number of years to maturity} and the \emph{Years since issuance} are selected by our robust regressions. Keep in mind these two variables are linked, via the maturity of the bond, thanks to the relation:  {\it Year to maturity} = {\it Maturity} - {\it Years since issuance}. Hence naturally, the coefficient of \emph{year to maturity} is negative while the one of \emph{years since issuance} is positive: the further away from the maturity, the smaller the bid-ask spread (in basis points).
  
      This could support the market folklore that there is only a short  time period  after the issuance when corporate bond trading is not too expensive  on secondary markets.
  \end{itemize}
  Other variables appearing in the OLS are not robust enough to be selected by penalized regressions. Removing these 17 variables from the regression only reduces the $R^2$ from around 0.55 to around 0.50, a minor price to pay for the increased robustness.  It is worth noticing that the $R^2$ of all these OLS and regularized regressions are around 50\%, which is in line with the best results obtained in the  literature: \cite{dn12bond} obtained $R^2$ between 0.50 and 0.80, while the $R^2$ of other studies were far below 0.50 (see Section \ref{sec:introduction}). The out-of-sample performance in Section \ref{sec:oos} further indicates the promise of the regression models.

In addition, the regression results (Table \ref{OLS_trace_comparison} for OLS and Table \ref{LASSO_standard_enhanced} for two-step LASSO) between the Enhanced TRACE  and  Standard TRACE datasets  confirm  similar performance in terms of $R^2$ for these two datasets. Therefore, one could choose either dataset for the bid-ask spread estimation, depending on the time-lag and the accuracy of the trade volume.

\section{Price Impact Analysis}
\label{sec:short}\label{sec:trade:by:trade} 
  After analyzing the average TCA on the weekly basis (Section \ref{sec:linreg}), we now move on to the second part of TCA analysis. We will focus on the individual  trade, and study  the amplitude of its price impact and the price impact decay after the transaction for liquid corporate bonds.  
  The goal of the price impact analysis is to determine the necessary set of events  to fit the mid-price dynamics and to understand how the impact of each type of event decays over time.

  As a benchmark comparison, recall  several stylized facts on equity market from \cite{bouchaud2009markets}:
 \begin{itemize}
     \item buy trades on average push the price up and sell trades on average drive the price down;
     \item the impact curve as a function of the volume of the trade is strongly concave. In other words, large volumes impact the price only marginally more than small volumes;
     \item the sign of market orders is strongly autocorrelated in time.
 \end{itemize}
To see whether these facts hold for corporate bonds, we 
apply a transient price impact model to estimate the price impact amplitude and decay pattern. Since the trading frequency is much lower on corporate bonds than on equities, it is more appropriate  to use the ``event time''  in our transient impact model instead of the chronological time used in other non-parametric price impact models for equity markets \cite{ citeulike:13675263,vker17omx}. 

Our first attempt is to model naively the mid-price by a single-event TIM (TIM1) (Section \ref{sec:TIM_single}). Using the signature plot as a metric for the goodness-of-fit  shows that TIM1 is not sufficient to describe the mid-price movements for corporate bonds. Meanwhile, statistical evidence implies an asymmetry between the price impacts from customer-buy orders and customer-sell orders, as detailed in Section \ref{sec:preliminary_asymmetry}. This statistical evidence motivates us to propose a TIM model with two types of events (TIM2) (Section \ref{sec:TIM_asymetric}): customer-buy orders and customer-sell orders.  The signature plot indicates  good performance of this improved TIM framework.

\subsection{Review: Transient Price Impact Models}\label{sec:methodology_tim}

Let us first review  the classic  transient impact model (TIM)  following  \cite{bouchaud2009markets} with the ``event time''. Assume  $\Pi$ is a set of event-types considered on the market and the mid-price $M^b_k$ of a corporate bond $b$ follows \cite{eisler2012price,taranto2018linear,lehalle2018market}:
\begin{eqnarray}\label{eq:tim_general}
M^b_k = \sum_{k' = k}^{-\infty} \sum_{\pi \in \Pi} \Big(G^b_\pi(k-k'){\bf} {\bf 1}(\pi^b_k=\pi)(V^b_{k'})^{\alpha}\epsilon^b_{k'}+\eta^b_{k'}\Big) + M^b _{-\infty},
\end{eqnarray}
where $\epsilon^b_k \in\{-1,+1\}$ is the sign of the $k$-th trade, estimation of which has been detailed in Section \ref{sec:RPT}, $V^b_k$ is the volume of the $k$th trade, $\pi^b_k$ is the type of the $k$-th trade, $\alpha$ is a power index, $G^b_\pi(\delta k)$ is a decaying kernel of type $\pi$ event, $\eta^b$ is a noise, and $M^b_{-\infty}$ is a initial value for the mid-price. $\eta$ is a random change of the fair price independent of $\epsilon$ and it is assumed to be i.i.d.

Note that $G^b_\pi$ is typically an exponential or a power law (i.e., $G^b_\pi(\delta t) 	\propto \exp (-\lambda \delta t)$ or $(1 + \delta t)^{-\gamma}$) \cite{eisler2012price,taranto2018linear,lehalle2018market}.
$G^b_\pi(\delta k)$ can be interpreted as the response function {\it per bond} when $\alpha=1$; $G_\pi^b(\delta k)$ can be understood as the response function {\it per order} when $\alpha=0$ and the volume is ignored. In equity markets, there has been empirical evidence showing that $\alpha \approx 0.1$ .

\subsection{First Attempt: Single-Event Transient Impact Model}\label{sec:TIM_single} 
In this section, we will show that the naive TIM1 model (i.e., Eqn \eqref{eq:tim_general} with one type of events) does not fit the price impact curves for cooperate bonds. 

To see this, note that the mid-price dynamics of bond $b$  under TIM1 are:
\begin{eqnarray}\label{eq:tim1}
M^b_k = \sum_{k' = k}^{-\infty} \Big(G^b(k-k')(V^{b}_{k'})^{\alpha}\epsilon^b_{k'}+\eta_{k'}^b\Big) + M^b _{-\infty},
\end{eqnarray}
where $\epsilon^b_k \in\{-1,+1\}$ is the sign of the $k$th trade as estimated in Section \ref{sec:RPT}, $V^b_k$ is the volume of the $k$th trade, $\alpha$ is a power index, $G^b(\delta k)$ is a decaying kernel of the bond $b$ mid-price, $\eta^b$ is noise and $M^b_{-\infty}$ is an initial value for the mid-price, and $\eta^b$ is a random change of the fair price independent of $\epsilon^b$ and is assumed to be i.i.d.

Under \eqref{eq:tim_general}, the change of the mid-price can be written as:
\begin{eqnarray}
R^b_k(1):= M^b_{k+1}-M^b_k = G^b(0)(V^b_{k+1})^{\alpha}\epsilon^b_{k+1} + \eta^b_{k+1} +\sum_{j = 0}^{\infty} \underbrace{(G^b(j +1) - G^b(j))}_{\Delta_1 G^b(j)}\cdot (V^{b}_{k-j})^{\alpha}\cdot \epsilon^b_{k-j}.
\end{eqnarray}
Consequently, we can check the values of ${S}^b(l) = \mathbb{E}\left[R^b_k(1)\epsilon^b_{k-l+1}\right]$ and ${C}^b(n) = \mathbb{E}\left[(V^b_{t+n})^{\alpha}\epsilon^b_{t+n}\epsilon^b_{t}\right]$ and obtain:
\begin{eqnarray}\label{eq:S_fun}
{S}^b(l) = G^b(0)C^b(l) +\sum_{j=0}^{+\infty}\Delta_1G^b(j)\cdot C^b(l-j-1).
\end{eqnarray}
If we only focus on the first $N$ transaction in the calculation of the response function, then \eqref{eq:S_fun} can be written in the following matrix format:
\begin{eqnarray}\label{eq:TIM_formula}
\underbrace{
\begin{pmatrix}
S^b(1)-G^b(0)C^b(1)\\
S^b(2)-G^b(0)C^b(2)\\
      \vdots\\
S^b(L)-G^b(0)C^b(L)
\end{pmatrix}}_{=:\overline{S}^b(L)}
\underbrace{
= \begin{bmatrix} 
C^b(0) & C^b(-1) & C^b(-2)&\cdots & C^b(-N+1)\\
C^b(1) & C^b(0) &C^b(-1) & \cdots & C^b(-N)\\
\vdots&\vdots&\vdots &\ddots& \vdots\\
C^b(L-1) &\cdots&\cdots&\cdots & C^b(L-N)
\end{bmatrix}}_{=:\overline{C}^b(N,L)}
\underbrace{
\begin{pmatrix}
\Delta_1G^b(0)\\
\Delta_1G^b(1)\\
      \vdots\\
\Delta_1G^b(N-1)\\
\end{pmatrix}}_{=:\overline{G}^b(N)}.
\end{eqnarray}
Note that it suffices to estimate  $S^b(l)$ and ${C}^b(n)$ for different values of $l$ and $n$, and the initial value $G^b(0)$. Afterwards, an estimator for $\overline{{G}}^b(N)$ can be constructed using \eqref{eq:TIM_formula} such that it follows:
\begin{eqnarray}\label{eq:TIM_inverse}
\widehat{\overline{{G}}^b(N)} = \widehat{\overline{{C}}^b(N,L)}^{-1}\cdot\widehat{\overline{S}^b(L)},
\end{eqnarray}
{with $\widehat{\overline{{C}}^b(N,L)}$ and $\widehat{\overline{S}^b(L)}$ the estimates of $\overline{C}^b(N,L)$ and $\overline{S}^b(L)$, respectively.}

To evaluate the model and quantify the price diffusion for different lags,  define the {\it signature plot} \cite{bouchaud2009markets} as below:
\begin{eqnarray}\label{eq:signature_plot}
D^b(l) = \frac{1}{l}\mathbb{E}\left[(M^b_{t+l}-M^b_t)^2\right].
\end{eqnarray}
For the TIM1 model, the approximated signature plot follows:
\begin{eqnarray}\label{eq:app_signature_plot1}
D^b_{{\rm TIM1}}(l) = \frac{1}{l} \sum_{0\leq n <l}\left(G^b(l-n)\right)^2 +\frac{1}{l}\sum_{n >0}\left(G^b(l+n)-G^b(n)\right)^2+2 \Phi^b(l)+D^b_{\rm const},
\end{eqnarray}
where  $D^b_{\rm const}$ is some constant and $\Phi^b(l)$ is the correlation-induced contribution to the price diffusion:
\begin{eqnarray*}
l \,\,\Phi^b(l) &=& \sum_{0 \leq n <m<l} G^b(l-n)G^b(l-m)C^b(m-n)\\
&+& \sum_{0 \leq n <m} [G^b(l+n)-G(n)][G^b(l+m)-G^b(m)]C^b(m-n)\\
&+&\sum_{0 \leq n <l}\sum_{m>0}G^b(l-n)[G^b(l+m)-G^b(m)]C^b(m+n).
\end{eqnarray*}


\paragraph{Experiment set-up.} {
We fit the TIM1 model \eqref{eq:TIM_formula}-\eqref{eq:TIM_inverse} with $L=N=10$ for the top-200 bonds (described in Section \ref{sec:price_imapct_bond_statistics}). In order to facilitate the comparison of different bonds, we calculate the propagator functions for relative price changes with $\alpha=0.0$.\footnote{The results with small $\alpha$ are qualitatively  the same.}}

\begin{figure}[ht]
\centering
        \begin{subfigure}[b]{0.4\textwidth}
                \centering
                \includegraphics[width=\textwidth]{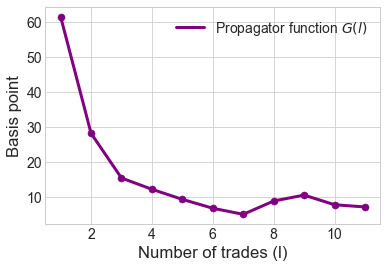}
                \caption{Propagator function $G$ for TIM1 model.}
                \label{fig:TIM1_propagator}
        \end{subfigure}%
        \begin{subfigure}[b]{0.4\textwidth}
                \centering
                \includegraphics[width=\textwidth]{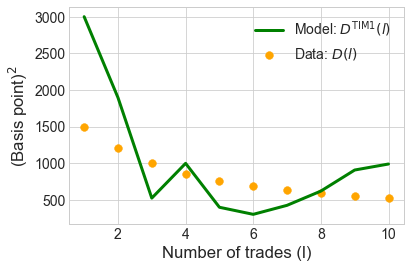}
               \caption{Signature plots: data vs. TIM1 model}
                \label{fig:TIM1_signature}
        \end{subfigure}%
\caption{Fitted TIM1 model and the goodness-of-fit (aggregation over 200 bonds and $\alpha=0$).}\label{fig:TIM1_result}
\end{figure}
From Figure \ref{fig:TIM1_result},  observe that:
\begin{itemize}
    \item unlike equity markets where the decay of the propagator function is {\it slow} for both large tick stocks and small tick stocks \cite{taranto2018linear}, the decay of the propagator functions is {\it fast} in cooperate bond markets and $G(l) \approx 10 (bp)$ when $l \ge 10$ (Figure \ref{fig:TIM1_propagator}).
    \item since the signature plot serves as a metric to evaluate the fitted models,  we observe from  Figure \ref{fig:TIM1_signature} that $D_{\rm TIM1}(l)$, the signature calculated from the TIM1 Model, does not fit well with the signature plot $D(l)$ calculated from the data. It appears that $D_{\rm TIM1}(l)$ overestimates the signatures for  small $l$.
\end{itemize}
\subsection{Modified TIM Model and Asymmetric Price Impact}\label{sec:asymmetry}

\subsubsection{Statistical Evidence of Asymmetric Price Impact}
\label{sec:preliminary_asymmetry}
 We next present some statistical evidence on the asymmetry of price impacts. This study  then motivates us to consider a two-type event model treating customer-buy and customer-sell orders separately.

To start, we adopt one-sided spread to test if there is any difference between the buy-side liquidity and the sell-side liquidity \cite{choi2019customer}:
\begin{eqnarray}\label{eq:one_sided_spread}
{\rm spread_{B}} &=& \frac{{\rm traded\,\, price} - {\rm reference\,\, price}}{ {\rm reference\,\, price}} \,\,{\bf 1}({\rm buy \,\, order}),\nonumber\\
{\rm spread_{S}} &=& \frac{{\rm reference \,\, price} - {\rm traded price} }{ {\rm reference\,\, price}} \,\,{\bf 1}({\rm sell \,\, order}).
\end{eqnarray}
For each customer trade, we calculate
its reference price as the volume-weighted average price of inter-dealer trades larger than \$100,000 in
the same bond-day, excluding inter-dealer trades executed within 15 minutes. ${\rm spread_{B}}$ and ${\rm spread_{S}}$ are calculated
at the bond-day level by taking the volume-weighted average of trade-level spreads.
The average buy spread is 44.52 (bp) and the average sell spread is 38.74 (bp) across all 1,993 liquid bonds. Afterwards, we perform a $t$-test with the null-hypothesis that the buy spread and the sell spread have the same sample mean. The null-hypothesis is rejected with a p-value smaller than 1\%, indicating that the buy spread is different from the sell spread. Meanwhile, we also perform $t$-tests for individual bonds – 1,215 out of 1,993 bonds have p-values smaller than 5\% indicating different buy spread and sell spread distributions.

\begin{figure}[H]  
\centering   
  \includegraphics[width=0.4\linewidth]{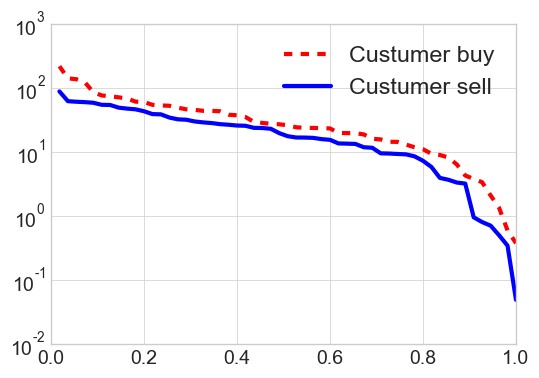}  
\caption{Rank frequency plot of buy-spread (${\rm spread_{B}}$) and sell-spread (${\rm spread_{S}}$) of all 1,993 bonds.}  \label{fig:rank_frequency}
\end{figure}  
The rank frequency plot of buy-spread ( ${\rm spread_{B}}$) and sell-spread (${\rm spread_{S}}$)  is visualized in Figure \ref{fig:rank_frequency}.
Note that a rank-frequency distribution is a discrete form of a quantile function (inverse cumulative distribution) in reverse order, giving the size of the element at a given rank. From Figure~\ref{fig:rank_frequency}, we observe that the distributions of the buy-spread and sell-spread have different tail behaviors.

\subsubsection{Modified TIM Models and Estimation of Asymmetric Price Impact}\label{sec:TIM_asymetric}

The preliminary analysis of price impact asymmetry from \ref{sec:preliminary_asymmetry} motivates us to propose a modified TIM model in which costumer-buy orders and costumer-sell orders are treated as different events in the calculation of propagator functions \cite{bouchaud2009markets,eisler2012price,eisler2016price,taranto2018linear,schneider2019cross}. See also \cite{jurksas2021liquidity} on liquidity spill-overs in sovereign bond market which estimate the price impact curves for buy and sell orders separately.

 This model is inspired by  \cite{taranto2018linear} where events with small trades and large trades are treated  differently.
Here we assume that there are two types of events $\Pi:=\{+1,-1\}$ with $+1$ denoting the customer-buy orders and $-1$ denoting the customer-sell orders. The calculation of the propagator function is similar to \eqref{eq:TIM_inverse} and as detailed in Appendix \ref{app:TIM2}. See also \cite[Appendix A.12]{lehalle2018market} for a more detailed discussion.

\paragraph{Experiment set-up.} {
We fit the TIM2 model \eqref{eq:TIM_formula}-\eqref{eq:TIM_inverse} with $L=N=10$ for the top-200 bonds (described in Section \ref{sec:price_imapct_bond_statistics}). Similar as before, we calculate the propagator functions for relative price changes in order to make different bonds comparable. In the estimation, we take $\alpha=0.0$ and the qualitative results for small $\alpha$ are similar.}

\begin{figure}[ht]
\centering
        \begin{subfigure}[t]{0.4\textwidth}
                \centering
                \includegraphics[width=\textwidth]{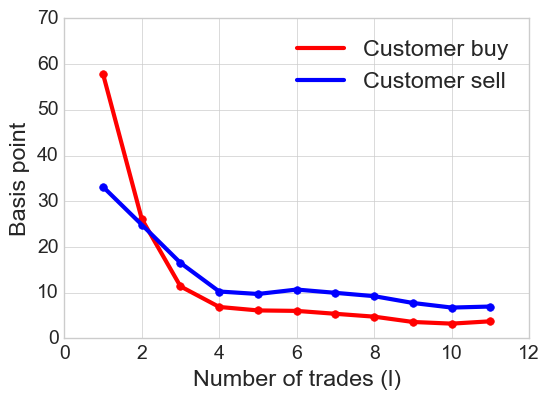}
                \caption{Propagator function $G_{+1}$ and $G_{-1}$ of the TIM2 model.}
                \label{fig:TIM2_propagator}
        \end{subfigure}%
        \begin{subfigure}[t]{0.4\textwidth}
                \centering
                \includegraphics[width=\textwidth]{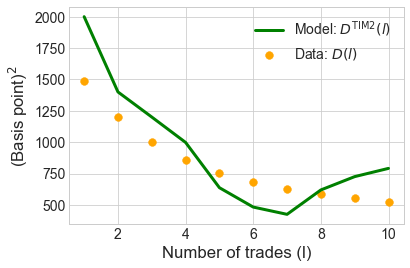}
               \caption{Signature plots: data vs. TIM2 model}
                \label{fig:TIM2_signature}
        \end{subfigure}%
\caption{Fitted TIM2 model and the goodness-of-fit ($\alpha=0$ and aggregation over 200 selected bonds).}\label{fig:TIM2_result}
\end{figure} 

One can observe the following from Figure \ref{fig:TIM2_result}: First, customer-buy orders have {\it larger} price impacts than customer-sell orders for the first few trade-times $l=1,2$ (Figure \ref{fig:TIM2_propagator}). Second, the decay of the propagator function for customer-buy orders are slightly {\it faster} than customer-sell orders (Figure \ref{fig:TIM2_propagator}). Moreover, comparing Figures \ref{fig:TIM1_signature} and \ref{fig:TIM2_signature}, we see the TIM2 model fits better with the signature plot calculated from the data. This implies that  the TIM2 model with customer-buy orders and customer-sell orders being treated differently is better than the single-event model TIM1.
\begin{figure}[t]
\centering
        \begin{subfigure}[b]{0.26\textwidth}
                \centering
                \includegraphics[width=\textwidth]{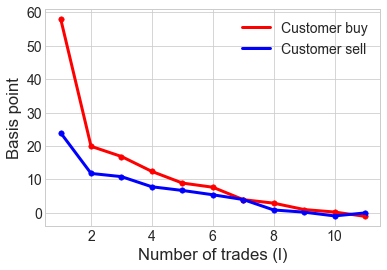}
                \caption{ Wells Fargo 94974BFY1 4.1\% (Issued 2 years and 10 years to maturity).}
                \label{fig:bond1}
        \end{subfigure}
        \hspace{1em}
        \begin{subfigure}[b]{0.26\textwidth}
                \centering
                \includegraphics[width=\textwidth]{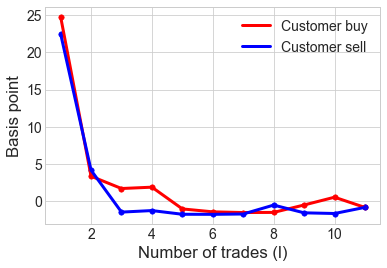}
               \caption{Goldman Sachs 38141GGQ1 5.25\% (Issued 5 years and 5 years to maturity).}
                \label{fig:TIM2_bond2}
        \end{subfigure}
        \hspace{1em}
        \begin{subfigure}[b]{0.26\textwidth}
                \centering
                \includegraphics[width=\textwidth]{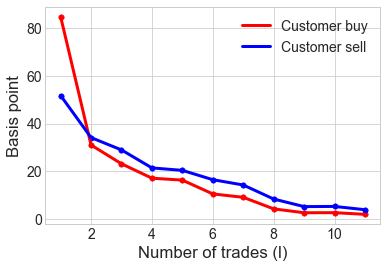}
               \caption{Transocean 893830AS8 6.0\% (Issued 9 years and 2 years to maturity).}
                \label{fig:TIM2_bond2}
        \end{subfigure}
\caption{Heterogeneity among bonds ($\alpha=0$).}\label{fig:TIM2_different_bonds}
\end{figure} 

 Figure \ref{fig:TIM2_different_bonds} suggests heterogeneity among bonds in terms of the  size of the market impact, the different impacts between buy-market orders and sell-market orders, and the shape of the decay. In addition, for newly issued bonds (i.e.,  Wells Fargo 94974BFY1 4.1\%) or bonds that are close to maturity (i.e., Transocean 893830AS8 6.0\%), the difference between the customer-buy propagator function and the customer-buy propagator function is {\it larger} than for the bonds that are in the middle of their life-time (i.e., Goldman Sachs 38141GGQ1 5.25\%).

It is worth pointing out that price impact models based on no-arbitrage considerations in equity markets require the price impact to be symmetric \cite{huberman2004price} and  \cite{gatheral2010no}. This no-arbitrage condition does not hold in bond markets due to the less liquidity  and more fragmentation in bond market. Our discovery of the asymmetric price impacts indicates  possible arbitrage opportunities in the secondary OTC market for corporate bonds.

 \paragraph{Summary.}   We propose to use propagator functions to measure the \emph{price impact of each single trade} for corporate bonds that are liquid enough. Our analysis finds two characteristics of the price impact of corporate bonds:
\begin{itemize}
\item The \emph{asymmetry between buying and selling trades}. The mid price moves triggered by a trade on a corporate bond are larger for buying transactions than those for selling ones.
  In terms of TCA, it means that the asset manager has to respect such an asymmetry and take it into consideration during the evaluation of the counterparty dealers.  
\item Decay in price impact curves, similar to the one identified in equity markets {\color{black} \cite{eisler2012price,taranto2018linear}}. The price impact curve consists of a jump corresponding to the adverse selection suffered by the dealer, followed by a decay stabilizing the price at the level of the permanent market impact.
\end{itemize}

\section{Conclusion}
This paper established a TCA benchmark in bond trading for retail investors and asset managers. 
It consists of (a) estimating the expected average cost  on a weekly basis via regularized regression analysis and (b) investigating   the amplitude of  price impact and the price impact decay  for each trade of liquid corporate bonds via TIM model.

The most important features identified in the regression analysis are volatility, trading activity, log(total volume), and issued years.   Meanwhile, asymmetry is discovered between buying and selling trades: mid-price moves triggered by a trade on corporate bonds are larger for buying  than those for selling.

Our study suggests the following  approach for TCA in practice:
\begin{enumerate}
\item For all corporate bonds of interest,  asset managers first compute an expected bid-ask spread given the characteristics of the bond and market conditions using one of the regression approaches proposed in Section \ref{sec:reglin}, and using either the Standard  TRACE  or the Enhanced  TRACE datasets for bid-ask spread approximation.

\item This reference bid-ask spread can be used to benchmark the bid-ask spread obtained while requesting  for quotes  from counterparties. It can also be used to \emph{score} all the obtained trades during the week. 
\item Worst trades can  be qualitatively evaluated using the average price impact curves obtained in Section \ref{sec:asymmetry}. More specifically, if a trade has  price impact larger than the curve showed in Figure \ref{fig:TIM2_different_bonds}, then it can be identified as a ``worst trade'' and the asset manager can conduct further analysis on the counterparty. 
\end{enumerate}


\begin{table}[H]
\centering\captionsetup{margin = 2.5cm}
\small{
\centering\setlength{\tabcolsep}{4pt}
\begin{tabular}{@{}l
*{3}{S[table-format=4.4,table-number-alignment=center]}@{}}\\ \toprule 
                       {}              &  {Estimate} &  {Standard error}&{t-value} \\ \midrule
\hline
Volatility                              &   77.7272{***}     &0.316&  151.687  \\
 Number of trade days                      &         -3.6648{***}   &   0.203  &  -18.014        \\

Prop number of buys                          &         10.3189{***}   &   0.689   &  14.971   \\
					                  
      Prop number of sells                           &      32.4532{***}  &    0.700  &   46.333      \\
                                                               
 {Trading activity}                               &    46.3169{***}   &   0.531  &   87.162  \\
                                                                  
        Prop volume sell \$                            &   16.3523{***}     & 0.608    & 26.893 \\
                                                   
        Prop volume buy \$                            &   26.4198{***}   &   0.642  &   41.155  \\
                                                                
 Log(total volume )                                   &    -21.4028{***}    &  0.272   & -78.757   \\
                                                              
        Avg price                                          &   -0.1175{***}  &    0.017  &   -6.723      \\
                                                     
      Coupon                                             &   -0.4707{***}   &   0.120    &    -3.914         \\
                                                                 
      Duration                                           &         1.4168{***}   &   0.148   &   9.596 \\
                                                           
      Years to maturity                               &       -0.0656   &   0.076   &  -0.858                   \\
                                                        
      Years since issuance                                   &        1.2552{***}    &  0.065  &   19.359      \\
                                                       
      Turnover                                          &    -1.7717    &  2.260   &  -0.784                  \\
                                                               
                                                              
       LIBOR-OIS                                     &      34.0088{***}    &  1.432    & 23.754               \\
                                                             
Indicator of high yield bonds                  &   26.7859{***}   &   0.595  &   45.039            \\
                                                               
Indicator of investment grade bonds      &  15.9861{***}   &   0.604   &  26.461             \\
                                                              
Indicator of basic materials sector         &  8.9009{***}   &   0.661  &   13.462                   \\
                                                             
Indicator of communications sector     &    5.1285{***}   &   0.374   &  13.695                 \\
                                                             
Indicator of consumer, cyclical sector    & 4.0261{***}   &   0.421   &   9.570               \\
                                                               
Indicator of consumer, non-cyclical sector   &4.9727{***}   &   0.378   &  13.140   \\
                                                              
Indicator of energy sector                 &    3.8824{***}   &   0.362  &   10.731             \\
                                                              
Indicator of financial sector                    &   4.0544{***}   &   0.325   &  12.460                      \\
                                                              
Indicator of industrial sector                  &   2.8149{***}    &  0.485    &  5.806                  \\
                                                               
Indicator of technology sector            & 4.4085{***}  &    0.489    &  9.017          \\
                                                               
Indicator of utilities sector                &   4.5836{***}   &   1.116   &   4.107           \\

Constant                                                 &   42.7720{***}    &  1.112   &  38.477        \\ \hline
N                                                             &   152,408                     &                  &                \\
$R^2$                                                     &    55.4\%                 &     &       \\
\bottomrule
\end{tabular}
\\ 
\scriptsize{Standard errors in parenthesis. Significance levels: * p$<$0.1, ** p$<$0.05, *** p$<$0.01. Two-tailed test. Source: TRACE Enhanced (2015-2016).} 
}
\caption{OLS regression: the impact  on bid-ask spread\label{OLS}}
\end{table}

\begin{table}[H]
\centering
\scriptsize{
\centering\setlength{\tabcolsep}{4pt}
\captionsetup{margin=-2.cm}
\begin{tabular}{@{}l
*{4}{S[table-format=4.4,table-number-alignment=center]}@{}}\toprule
                    & { Model OLS } &   {Model L1} & {Model L2} & {Model L3}  \\
                    &  {All features}  &   {$\lambda_l = 1.13$ }&{ $\lambda_l = 2.98$ }&{ $\lambda_l = 7.85$} \\
                     &  {(Benchmark)}  &  & &  \\\midrule
Volatility        &      77.7272{***}      &      80.7218{***}         &       83.5410{***}             & 86.5025{***}\\
                  &    {  (0.316)    }      &        {(0.253)      }      &     { (0.221)    }           &{(0.237)} \\
Number of trade days                            &        -3.6648 {***}     &                                  &                                  &        \\
                                                                   &      {( 0.203)}          &                                &                                 &    \\
    Prop number of buys                          &        10.3189{***}     &                 &                                 & \\
					                         &       { (0.689)     }      &                     &                                    & \\
      Prop number of sells                           &        32.4532{***}      &              &                                &  \\
                                                                  &      {(0.700)  }         &                    &                              & \\ 
 {Trading activity}                              &    46.3169{***}         &                       42.1450{***}        &    39.0703{***}                &   \\
                                                                  &      {(0.531)}          &            { (0.451)  }    &     { (0.433)  }    &    \\
        Prop volume sell \$                              &16.3523{***}            &                                 &                                &        \\
                                                                  &     { (0.608) }         &                                  &                                &         \\
        Prop Volume buy \$                              &  26.4198{***}          &                                   &                                &        \\
                                                                  &     { (0.642)    }       &                                  &                                 &         \\
 Log(total volume )                                   & -21.4028{***}            &        -21.1246  {***}       &       -19.4449{***}              &        \\
                                                                  &      {(0.272)   }        &         {(0.252)  }         &      {   (0.243)  }        &         \\
        Avg price                                          &      -0.1175 {***}         &               &                & \\
                                                                 &     { (0.017)}           &                   &             &  \\
      Coupon                                             &     -0.4707 {***}         &  1.2071{***}                                   &                         & \\
                                                                 &     { (0.120)}            &     {(0.089)}                              &                             &\\
      Duration                                           &   1.4168 {***}           &          0.2453{***}         &            & \\
                                                                &     { (0.148) }           &         {(0.035) }           &               & \\
      Years to maturity                               &        -0.0643  {}         &        -0.1078           &                              &    \\
                                                                &    {  (0.076) }           &          {(0.075)}             &                              &    \\
      Years since issuance                                   &     1.2576{***}           &        -0.0216             &        0.2336{***}       &  1.0396 {***} \\
                                                                &     { (0.065)  }          &          {(0.052)  }           &      {   (0.046) }        &{ (0.011) }  \\
      Turnover                                          &       -1.7717{}          &               &                              &    \\
                                                               &   {   (2.260)}              &                 &                               &    \\
                                                               &     { (1.542)}             &                    &                                 & \\
       LIBOR-OIS                                     &     34.0088{***}                 &                &                               &  \\
                                                               &    {  (1.432) }             &                     &                                &   \\
Indicator of high yield bonds                  &     26.7859{***}                 &          &                               &  \\
                                                               &      {( 0.595)}          &               &                              &   \\
Indicator of investment grade bonds      &   15.9861{***}               &        &                               &   \\
                                                               &      {(0.604)}               &                                  &                              &    \\
Indicator of basic materials sector         &    8.9009{***}                 &                                 &                               &  \\
                                                               &   {(0.661) }               &                                   &                              &   \\
Indicator of communications sector     &    5.1285{***}                   &                                  &                               &  \\
                                                              &   {(0.374  ) }                  &                                  &                              &   \\
Indicator of consumer, cyclical sector    &    4.0261{***}                &                                  &                               &  \\
                                                               &   {(0.421 )}                  &                                   &                              &   \\
Indicator of consumer, non-cyclical sector   &    4.9727 {***}       &                                    &                               &  \\
                                                               &   {(0.378) }                &                                   &                              &   \\
Indicator of energy sector                       &     3.8824{***}           &                                  &                               &  \\
                                                               &  { (0.362)  }               &                                  &                              &   \\
Indicator of financial sector                    &    4.0544{***}              &                                 &                               &  \\
                                                               &   {(0.325  ) }                 &                               &                              &   \\
Indicator of industrial sector                  &    2.8149 {***}  &                               &                               &  \\
                                                                &   {(0.485) }                 &                               &                              &   \\
Indicator of technology sector            &    4.4085{***}                   &                                &                               &  \\
                                                               &  { (0.489)  }                &                                &                              &   \\
Indicator of utilities sector                &     4.5836 {***}                   &                                    &                               &  \\
                                                               &  { (1.116 )}                &                                  &                              &   \\
                                                               
Constant                                                 &      42.7720             &    86.1737          &      85.6982            &-2.9270\\ \hline
N                                                             &   152,408                     &    152,408              &   152,408                &152,408 \\
$R^2$           &    55.4\%                 &      52.8\%            &       51.5\%         &43.22\%  \\ \bottomrule
\hline
\end{tabular}
\\ 
\scriptsize{Standard errors in parenthesis. Significance levels: * p$<$0.1, ** p$<$0.05, *** p$<$0.01. Two-tailed test. Source: TRACE Enhanced (2015-2016).} 
}
\caption{Two-step Lasso regression table: the impact  on bid-ask spread (in bp)\label{LASSO_normal_scale}}
\end{table}


\begin{table}[H]
\scriptsize{
\centering \setlength{\tabcolsep}{4pt}
\captionsetup{margin=2.5cm}
\begin{tabular}{@{}l
*{4}{S[table-format=2.4,table-number-alignment=center]}@{}}\toprule
                    &  {Model OLS } &  { Model R1} & {Model R2}& {Model R3}  \\
                    &  {All features}  &   {$\lambda_r=1.62\times 10^{4}$} & {$\lambda_r=6.95\times 10^{4}$} & {$\lambda_r=1.27\times 10^{6}$} \\
                     &  {(Benchmark)}  &  & &\\
\midrule
Volatility                                               &       77.7272{***}          &   82.0769{***}        &    84.0327{***}         &75.1199 {***}\\
                                                         &      {(0.316)}             &     {(0.522)}        &      {(0.364)}          & {(0.380)} \\
Number of trade days                                     &         -3.6648{***}       &   1.1389 {***}       &   1.2833{***}           & 0.1705       \\
                                                         &     { ( 0.202) }           &    {(0.339)}         &    {(0.241) }           &   {(0.249) }            \\
    Prop number of buys                                  &        10.3189{***}        &  -3.5702{***}        &  -1.1382                & -0.0734 \\
					                                     &        {(0.689)}           &    {(1.148) }        &     {( 0.817) }         & {(0.844 )} \\
      Prop number of sells                               &        32.4532{***}        &    3.5702{***}       &    1.1382               &0.0734 \\
                                                         &      {(0.700)}             &   {(1.167)}          &       {(0.830)}         &{(0.858)} \\ 
  {Trading activity}                                  &     46.3169{***}           &    14.2859{***}      &     4.3609{***}         & 0.2794 \\
                                                         &      {(0.531)}             &      {(0.885)}       &     {(0.630 ) }          &  {(0.651) } \\
        Prop volume sell \$                                & 16.3523***                 &       -0.7126        &       -0.0237           &      0.0060   \\
                                                         &    {  (0.608)  }           &       {(1.013)}      &       {(0.720) }        & {(0.745)    }    \\
        Prop volume buy \$                                 &  26.4198{***}              &        0.7126        &      0.0237             &     -0.0060   \\
                                                         &     { (0.642)    }         &        {(1.069)}     &        {(0.761)  }      &   { (0.786) }      \\
 Log(total volume )                                      &  -21.4028{***}             &     -10.5886{***}    &       -4.5781{***}      &   -0.4096     \\
                                                         &    {  (0.272)   }          &       {(0.453)}      &     {  (0.322 )  }      &     { (0.333 ) }  \\
        Avg price                                        &     -0.1175{***}           &   -0.1313{***}       &     -0.0896{***}        &  -0.2174 {***} \\
                                                         &     { (0.017)  }           &       {(0.029) }     &    { (0.021)  }         &{ (0.021)} \\
      Coupon                                             &     -0.4707{***}           &    0.0114            &     -0.0264             &0.2436 \\
                                                         &     { (0.120)    }         &      {(0.200) }      & { (0.142)  }            & {(0.147)}\\
      Duration                                           &     1.4168{***}            &       1.7553{***}    &   0.0044{***}           & 0.2660       \\
                                                         &     { (0.1480)  }          &      {(0.246)}       &  {(0.003)}              & {(0.181)} \\
      Years to maturity                                   &       -0.0643              &        -0.6328 {***} &    -0.5433{***}         &  0.0510 \\
                                                         &      {(0.076) }            &         {(0.127) }   &    {( 0.091)}           &{ (0.094)  } \\
      Years since issuance                                       &     1.2576{***}            &      1.0926{***}     &      1.0705 {***}       & 0.6555{***} \\
                                                         &     { (0.065)   }          &      {(0.108)}       &      { (0.077)  }       &    {(0.079) }\\
      Turnover                                           &       -1.7717              &     -0.0963          &      -0.0893      &    -0.0120  \\
                                                         &     { (2.260)   }          &   {(3.764)}          &    {(2.677)}            &  {(2.768)   }   \\
                                                         &     { (0.012) }            &  {(0.002) }          &  { (0.029)}             & {(0.018)}  \\
       LIBOR-OIS                                         &     34.008{***}            &          2.5574      &     0.6844              & 0.0386  \\
                                                         &    {  (1.432)  }           &      {(2.384)}       &     {(1.696)}           & {(1.753)}  \\
Indicator of high yield bonds                            &    26.7859{***}            &        1.8302{*}     &       0.5512            &  0.0478    \\
                                                         &  { (0.595)  }              &    {(0.990)}         &        { ( 0.716) }     & {(0.728) }    \\
Indicator of investment grade bonds                      &   15.9861{***}             &     -1.8302{*}       &    -0.5512              & -0.0478  \\
                                                         &    {  (0.604)   }          &   {(1.006) }         &     {   (0.012)   }     & { (0.740) }       \\
Indicator of basic materials sector                      &     8.9009{***}            &      0.6401          &  0.1724                 & 0.0125\\
                                                          &   {(0.661)    }           &      {(1.101)}       &       {  (0.783)}       & { (0.810) } \\
Indicator of communications sector                       &    5.1285{***}             &       0.7563         & 0.3777                  &0.0373  \\
                                                         & {  (0.374)     }           & {(0.624) }           &     {   (0.444)   }     &  {(0.459 ) }  \\
Indicator of consumer, cyclical sector                   &    4.0261{***  }           &      -0.5567         &    -0.2340              &  -0.0256    \\
                                                         &  { (0.421)  }              &   {(0.701)  }        &   {  (0.498 )}          & {(0.515 )  }   \\
Indicator of consumer, non-cyclical sector               &    4.9727{***}             &       -0.4812        &       -0.2448           & -0.0371 \\
                                                         & {  (0.378)  }              &     {(0.630) }       &     {  (0.448  ) }      &{  (0.463 )  } \\
Indicator of energy sector                               &   3.882{***}               &    -1.3953{**}       &       -0.7044{*}        &   -0.0304 \\
                                                         &  { (0.362) }               &    {(0.603) }        &   {    (0.429 ) }       &  {(0.433) }  \\
Indicator of financial sector                            &     4.0544{***}            &      0.1134          &     0.0911              &   -0.0054 \\
                                                         &  { (0.325 )   }            &   {(0.542)}          &        { (0.385) }      &   { (0.398)}\\
Indicator of industrial sector                           &    2.8149{***}             &      -0.1199         &    0.0657               &0.0130 \\
                                                         & {  (0.485)  }              &  {(0.807)}           &           { (0.574)   } &{(0.594 )}    \\
Indicator of technology sector                           &    4.4085{***}             &    1.0098            &       0.4649            &0.0346   \\
                                                         &   {(0.489 )    }           &    {(0.814) }        &     {   (0.579 ) }      &{ (0.599)}   \\
Indicator of utilities sector                            &    4.5836{***}             &    0.0335            &       0.0114            &0.001   \\
                                                         &  { (1.116)     }           &     {(1.859)}        &    {   (1.322) }        & { (1.367)} \\
                                                               
Constant                                                 &        42.7720             &        0.000         &   0.000                 &0.000 \\ \hline
N                                                        &   152,408                  &    152,408           &  152,408                &152,408 \\
$R^2$                                                    &    55.4\%                  &      53.5\%          &      52.0\%             &50.0\%  \\
\hline
\end{tabular}
\\
\scriptsize{Standard errors in parenthesis. Significance levels: * p$<$0.1, ** p$<$0.05, *** p$<$0.01. Two-tailed test. Source: TRACE Enhanced (2015-2016).} 
}
\caption{Ridge regression table: the impact on bid-ask spread (in bp)\label{Ridge_normal_scale}}
\end{table}


%

\begin{table}[H]
\scriptsize{
\centering\setlength{\tabcolsep}{4pt}
\captionsetup{margin=-0.5cm}
\begin{tabular}{@{}l
*{4}{S[table-format=4.4,table-number-alignment=center]}@{}}\toprule
                    & { Model OLS } &   {Model EN1} & {Model EN2} & {Model EN3}  \\
                    &  {All features}  &   {$(\alpha,\lambda_e) = (0.5,0.774)$ }&{ $(\alpha,\lambda_e) = (0.8,2.15)$ }&{ $(\alpha,\lambda_e) =(0.5,129)$} \\ 
                       &  {(Benchmark)}  &  & &\\\midrule
Volatility        &     77.7272{***}      &     77.8015{***}         &       83.5410{***}             &   87.5398{***}\\
                  &    {  (0.316)    }      &        {(0.300)      }      &     { (0.221)    }           &{(0.253)} \\
Number of trade days                            &        -3.6648 {***}     &     -3.3792{***}                              &                                  &        \\
                                                                   &      {( 0.203)}          &           {( 0.200)}                      &                                 &    \\
    Prop number of buys                          &        10.3189{***}     &         20.4300  {***}      &                                 & \\
					                         &       { (0.689)     }      &       { (0.725)     }                 &                                    & \\
      Prop number of sells                           &        32.4532{***}      &      36.5251        &                                &  \\
                                                                  &      {(0.700)  }         &      {(0.732)  }               &                              & \\ 
 {Trading activity}                              &    46.3169{***}         &                       45.9632{***}        &    39.0703{***}                &   \\
                                                                  &      {(0.531)}          &            { (0.519)  }    &     { (0.433)  }    &    \\
        Prop volume sell                               &16.3523{***}            &                                 &                                &        \\
                                                                  &     { (0.608) }         &                                  &                                &         \\
        Prop Volume buy                               &  26.4198{***}          &                                   &                                &        \\
                                                                  &     { (0.642)    }       &                                  &                                 &         \\
 Log(total volume )                                   & -21.4028{***}            &        -21.6955 {***}       &       -19.4449{***}              &        \\
                                                                  &      {(0.272)   }        &         {(0.252)  }         &      {   (0.243)  }        &         \\
        Avg price                                          &      -0.1175 {***}         &     -0.1083 {***}          &                & 0.1264{***} \\
                                                                 &     { (0.017)}           &       {(0.016)}            &             & {(0.014)} \\
      Coupon                                             &     -0.4707 {***}         &                                    &                         & \\
                                                                 &     { (0.120)}            &                                   &                             &\\
      Duration                                           &   1.4168 {***}           &          1.5875{***}         &            & \\
                                                                &     { (0.148) }           &         {(0.146) }           &               & \\
      Years to maturity                               &        -0.0643  {}         &        -0.1680           &                              &    \\
                                                                &    {  (0.076) }           &          {(0.074)}             &                              &    \\
      Years since issuance                                   &     1.2576{***}           &        0.9346{***}             &        0.2336{***}       &  \\
                                                                &     { (0.065)  }          &          {(0.056)  }           &      {   (0.046) }        &  \\
      Turnover                                          &       -1.7717{}          &               &                              &    \\
                                                               &   {   (2.260)}              &                 &                               &    \\
                                                               &     { (1.542)}             &                    &                                 & \\
       LIBOR-OIS                                     &     34.0088{***}                 &                &                               &  \\
                                                               &    {  (1.432) }             &                     &                                &   \\
Indicator of high yield bonds                  &     26.7859{***}                 &     33.1587{***}      &                               &  \\
                                                               &      {( 0.595)}          &    {( 0.650)}           &                              &   \\
Indicator of investment grade bonds      &   15.9861{***}               &   23.7964{***}      &                               &   \\
                                                               &      {(0.604)}               &    {(0.625)}                              &                              &    \\
Indicator of basic materials sector         &    8.9009{***}                 &                                 &                               &  \\
                                                               &   {(0.661) }               &                                   &                              &   \\
Indicator of communications sector     &    5.1285{***}                   &                                  &                               &  \\
                                                              &   {(0.374  ) }                  &                                  &                              &   \\
Indicator of consumer, cyclical sector    &    4.0261{***}                &                                  &                               &  \\
                                                               &   {(0.421 )}                  &                                   &                              &   \\
Indicator of consumer, non-cyclical sector   &    4.9727 {***}       &                                    &                               &  \\
                                                               &   {(0.378) }                &                                   &                              &   \\
Indicator of energy sector                       &     3.8824{***}           &                                  &                               &  \\
                                                               &  { (0.362)  }               &                                  &                              &   \\
Indicator of financial sector                    &    4.0544{***}              &                                 &                               &  \\
                                                               &   {(0.325  ) }                 &                               &                              &   \\
Indicator of industrial sector                  &    2.8149 {***}  &                               &                               &  \\
                                                                &   {(0.485) }                 &                               &                              &   \\
Indicator of technology sector            &    4.4085{***}                   &                                &                               &  \\
                                                               &  { (0.489)  }                &                                &                              &   \\
Indicator of utilities sector                &     4.5836 {***}                   &                                    &                               &  \\
                                                               &  { (1.116 )}                &                                  &                              &   \\
                                                               
Constant                                                 &      42.7720             &    56.9551          &      85.6982            &-12.7213\\ \hline
N                                                             &   152,408                     &    152,408              &   152,408                &152,408 \\
$R^2$           &    55.4\%                 &      53.8\%            &       51.5\%         &42.62\%  \\ \bottomrule
\hline
\end{tabular}
\\ 
\scriptsize{Standard errors in parenthesis. Significance levels: * p$<$0.1, ** p$<$0.05, *** p$<$0.01. Two-tailed test. Source: TRACE Enhanced (2015-2016).} }
\caption{Two-step EN regression table: the impact  on bid-ask spread (in bp)\label{Enet_normal_scale}}
\end{table}

\newpage
\bibliographystyle{apalike}
\bibliography{lehalle_bkbonds.bib}
\newpage
\appendix
\section{Literature review}
\label{sec:literature2}
\begin{table}[!h]
  \centering
  \begin{tabular}{llc}
Reference & Dataset(s) Name(s) & Period covered\\\hline
\cite{schu01bonds} & 
CAI & 1995-1997\\
\cite{chakravarty2003trading} & 
NAIC & 1995-1997\\
\cite{citeulike:75856} & 
TAQ, ISSM (Nyse),  GovPX & 1991-1998 \\
\cite{bessembinder2006market} & 
NAIC + TRACE & 2001 ; 2002\\
\cite{gold07bond} & 
TRACE & 2002-2004\\
\cite{citeulike:11289424} & %
Nyse Archives & 1926-1930; 1943-1948\\
\cite{dn12bond} & 
TRACE & 2003-2005\\
\cite{asquith2013market} & 
TRACE & 2004-2007\\
\cite{citeulike:13794923} & 
 Labelled TRACE & 2003-2013\\
\cite{finra15bonds} & 
TRACE & 2002-2007\\
\cite{hender11bonds} & 
TRACE + MarketAxes & 2011-2011\\
\cite{ales16bond}& 
TRACE & 2004-2012
    \\\hline
  \end{tabular}
  \caption{List of empirical papers on transaction costs of corporate bonds}
  \label{tab:comp:literature}
\end{table}

\section{Data Processing}
\subsection{Assigning a sign to a trade and identifying RPT}
\label{sec:signtrades}

To estimate the sign of transactions, we will first reproduce the essentials of  preprocessing to identify such RPTs in \cite{citeulike:13771761}.
 We identify potential RPTs as pairs of sequentially adjacent trades of the same size for which one trade is a customer trade. To find these trades in the Enhanced TRACE data, we first identify all size runs (sequences) of two or more trades of equal 
 size. Next, for each size run, we consider which trades, if any, consist of a pair of trades in a potential RPT. We identify potential RPTs if one trade of two adjacent trades within a size run is a dealer trade with a customer, or if both trades in an adjacent pair are customer trades {\it and} the dealer both buys and sells. We identify the first such pair as a potential RPT, and then continue searching the size run for any additional pairs that do not involve trades already identified as being part of a potential RPT. 
%
\cite{citeulike:13771761} found that the RPT rate is above 42\% and 41\% of customer trades appear to be RPTs. The RPT rate for our entire Enhanced TRACE data set is 23.9\%. Moreover, Table \ref{tbl:RPT-1993} shows we found 21.8\% RPTs.

\begin{table}[H]
\centering\captionsetup{margin = 2.5cm}
 \setlength{\tabcolsep}{8pt}
\begin{center}
\begin{tabular}{@{}l
*{3}{S[table-format=5.3,table-number-alignment=center]}@{}}\toprule
{}&{Total}&{Dealer-customer}&{Dealer-dealer}\\ \midrule
    Total number& 9,413,109&4,523,268&4,889,841\\
    Number of RPT&2,052,644&1,145,127&907,517\\
    Percentage of RPT &{21.8{\%}}&{25.3{\%}}&{18.5{\%}}\\ \bottomrule
\end{tabular}
\subcaption{Statistics of potential RPTs for selected 1,993 bonds.}
\label{tbl:RPT-1993} 

\begin{tabular}{@{}l
*{3}{S[table-format=5.3,table-number-alignment=center]}@{}}\toprule
{}&{Total}&{Customer-buy}&{Customer-sell}\\ \midrule
       Total number& 3,378,141&1921608&1456533\\
     Number percentage& {100\%}&{57\%}&{43\%}\\
    Total volume&{$3.10\times 10^{12}$}&{$1.73\times 10^{12}$}&{$1.37\times 10^{12}$}\\
   Volume percentage &{100\%}&{55.8\%}&{44.2\%}\\ \bottomrule
\end{tabular}
\subcaption{Statistics of non-RPT dealer-customer trades.}
\label{tbl:nonRPT-1993}
\end{center}
\caption{Statistics of selected 1,993 bonds for the BA-spread regression.}\label{1993bonds}

\end{table}

\begin{table}[H]
\centering\captionsetup{margin = 2.5cm}
 \setlength{\tabcolsep}{8pt}
\begin{center} 
\begin{tabular}{@{}l
*{3}{S[table-format=5.3,table-number-alignment=center]}@{}}\toprule
{}&{Total}&{Dealer-customer}&{Dealer-dealer}\\ \midrule
    Total number& 3,251,042&1,387,290&1,819,731\\
    Number of RPT&783,022&414,503&365,459\\
    Percentage of RPT &{24.1{\%}}&{29.9{\%}}&{20.1{\%}}\\ \bottomrule
\end{tabular}
\subcaption{Distribution of potential RPTs.}
\label{tbl:RPT}

\begin{tabular}{@{}l
*{3}{S[table-format=5.3,table-number-alignment=center]}@{}}\toprule
{}&{Total}&{Customer-buy}&{Customer-sell}\\ \midrule
    Total number&972,787&566,232&406,555\\
     Number percentage& {100\%}&{58\%}&{42\%}\\
    Total volume&{$5.12\times 10^{11}$} & {$2.68\times 10^{11}$} & {$2.45\times 10^{11}$}\\
   Volume percentage &{100\%}&{52.3\%}&{47.7\%}\\ \bottomrule
\end{tabular}
\subcaption{Distribution of  non-RPT Dealer-customer trades.}
\label{tbl:nonRPT}
\end{center}
\caption{Statistics of selected {\color{black}200} bonds for the price impact analysis.}\label{1000bonds}
\end{table}

\subsection{Data Filtering}\label{filtering}
The data cleaning procedure combines the approaches in \cite{DICK2014} and \cite{citeulike:13771761}, with  the following steps: 
\begin{enumerate}
\item Remove canceled trades and apply corrections to ensure that only trades that are actually settled are accounted for. After the removal of canceled trades and  canceled corrections records, there are 32,931,539 trades.

\item Remove the transactions reported by agents as  both principal and agent in the dealer-to-dealer transactions report  to FINRA  (see \cite{DICK2014}). As a result,  2,095,934 (6.36\%) of the reports are removed, with 30,835,605 reports remaining after this step.

\item Remove the transactions on unusual trading days  such as weekends and holidays. Thus 
5,753 (0.02\%) records are removed, with  30,829,852 reports left after this step.

\item Exclude all trade reports with an execution time outside of the normal 8:00AM to 5:15PM ET trading hours. Therefore 745,619 (2.4\%) are removed, with  30,084,233 reports remaining after this step. 

\item Remove all irregular trades with sales condition codes that indicate late reports, late reports after market hours, weighted average price trades, or trades with special price flags. As a result, 583,157 (1.9\%) reports are removed, with  29,501,076 reports left after this step.

\item Remove trade reports with a price below 10. This price filter step excludes 217,321 (0.74\%) of the remaining trades, with 29,283,755 reports left after this step.

\item Select reports classified as {\it corporate bonds} in the dataset. Remove those reports with sub-product indicators such as Mortgage Backed Securities Transactions. Consequently 563,942  (1.94\%) of the remaining reports are filtered out, with 28,719,813 reports  left.
\end{enumerate}

\begin{table}[ht]\small{
    \centering\captionsetup{margin = 1.5cm}
    \setlength{\tabcolsep}{8pt}
    \captionsetup{margin=2.5cm}
    \begin{center}
      \begin{tabular}{lrrr}
        &Removal (nbe)& Removal (pct)& Number left\\\toprule
        Step 1\\
        Keep settled trades &1,877,866 &{5.4\%} &32,931,539\\\midrule
        Step 2\\
        Keep trades reported by dealers &2,095,934 & {6.36\%}&30,835,605\\\midrule
        Step 3\\
        Keep business days  &5,735 &{0.02\%}&30,829,853\\\midrule
        Step 4\\
        Keep opened hours  &745,619 &{2.4\%}& 30,084,233\\\midrule
        Step 5\\
         Keep regular trades           &583,157 &{1.9\%}&29,501,076\\\midrule
        Step 6\\
        Keep compatible prices                              &217,321 &{0.074\%}&29,182,755\\\midrule
        Step 7\\
        Keep bonds only                    &563,942 &{1.94\%}&28,719,813\\\midrule
        Selection for LR\\
        For bid-ask spread regression & -- & -- & 4,371,363\\\midrule
       Selection for PI \\
        For price impact curves & -- &-- & 1,404,507\\
        \bottomrule
      \end{tabular}
    \end{center}
            }
             \caption{ \label{tbl:data_clean} Data filtering procedure.}
\end{table}

\section{Statistical Test and Regression Analysis}
\subsection{ANOVA test and Kruskal-Wallis H-test}\label{a_h_tests}

Suppose there are $W$ groups of observations. (In our example, $W=6$.) There are $n_w$ observations in group $w$ and the total number among all groups is $n$. Within each group, $w=1,2,\cdots,W$ and the observations are denoted as $y_{w,1},\cdots,y_{w,n_w}$ with sample size $n_w$. Denote $\bar{y}_w=\frac{\sum_{i=1}^{n_w}y_{w,i}}{n_w}$ as the sample mean in group $w$ and $\bar{y}=\frac{\sum_{w=1}^{W} \sum_{i=1}^{n_w}y_{w,i}}{n}$ as the sample mean of all observations.

\paragraph{One-way ANOVA test.}
A one-way ANOVA test is applied to samples from two or more groups, possibly with differing sizes. In a one-way ANOVA test, the formula for the F-ratio is $F = \frac{MS_B}{MS_W}$,
where $MS_B=\frac{\sum_{j=1}^J n_j(\bar{y}_j-\bar{y})^2}{n-1}$ is the between-group mean square value and $MS_W = \frac{\sum_{j=1}^J \sum_{i=1}^{n_j}(y_{j,i}-\bar{y}_j)^2}{\frac{n(n-1)}{W}}$ is the within-group mean square value.


\subsection{KS test} \label{KS_test}
Denote by $F(x)=\mathbb{P}(X_1 \leq x)$ a cumulative density function of a true underlying distribution of the data and  define an {\it empirical} cumulative density function by
$F_n(x) =\mathbb{P}_n (X \leq x) = \frac{1}{n} \sum_{i=1}^n I(X_i \leq x).$
Then $
\sup_{x \in \mathbb{R}} |F_n(x)-F(x)| \rightarrow 0.$

Suppose that the first sample $X_1,X_2,\cdots,X_m$ of size $m$ has a cumulative distribution function (CDF) $F(x)$ and the second sample $Y_1,Y_2,\cdots,Y_n$ of
size $n$ has a CDF $G(x)$. Suppose that one wants to test \[H_0: F=G\,\,\,\,\, \mbox{vs.} \,\,\,\,\, H_1: F \neq G.\]
Let $F_m(x)$ and $G_n(x)$ be their respective empirical CDFs,  then 
$
D_{mn} = \left(\frac{mn}{(m+n)}\right)^{\frac{1}{n}} \sup_{x}\left|F_m(x)-G_n(x)\right|
$
satisfies the following property of convergence in the distribution:
\[
\mathbb{P}\left( |D_{mn}|<t \right) \rightarrow H(t) =1-2\sum_{i=1}^{\infty}(-1)^{i-1}e^{-2i^2t},
\]
where $H(t)$ is the CDF of the KS distribution.

\paragraph{Kruskal-Wallis H-test.}
The Kruskal-Wallis H-test is a non-parametric version of ANOVA. The test works on two or more independent samples, which may have different sizes. 
The mathematical formula for H-statistic is $$
H = \frac{12}{n(n+1)} \sum_{j=1}^W \frac{T_j^2}{n_j} -3(n+1),$$
where $T_j$ is the sum of ranks in the $j^{th}$ group.

\subsection{Cross-validation results}\label{cross_validation}
\subsection*{Lasso.}
\begin{figure}[H]
  \begin{center}
    \includegraphics[width=0.6\columnwidth]{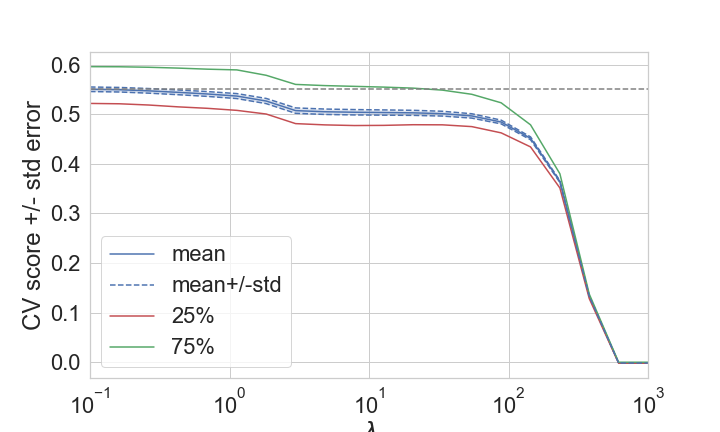}
    \caption{\label{fig:cross_validation} Cross-validation score for Lasso.}
  \end{center}
\end{figure}
\begin{table}[H]
  \centering\captionsetup{margin = 2.5cm}
  \setlength{\tabcolsep}{8pt}
\begin{tabular}{||c| c c c c c ||} 
 \hline
$\lambda_l$ & {$1.0\times 10^{-1}$} & {$1.62\times 10^{-1}$} & {$2.64\times 10^{-1}$} &  {$4.28\times 10^{-1}$} & {$6.95\times 10^{-1}$} \\ [0.5ex] 
 \hline\hline
$\widehat{R^2}$ &  0.550&  0.549&  0.547&  0.544&  0.541\\
Number in $\frak{I}_1$ &   15& 12& 12& 14& 15\\
Number in $\frak{I}_2$ & 152& 153& 155& 156& 157  \\
 \hline
  \hline
$\lambda_l$ &1.13& 1.83& 2.98& 4.83& 7.85\\ [0.5ex] 
 \hline\hline
 $\widehat{R^2}$ &  0.527&  0.517&  0.507&  0.505&  0.484\\
Number in $\frak{I}_1$ & 18& 16& 16& 16& 14\\
Number in $\frak{I}_2$  & 158& 160& 161& 161& 162  \\
 \hline
  \hline
$\lambda_l$ &  {$1.27\times 10$}  & {$2.07\times 10$} & {$3.36\times 10$} & {$5.46\times 10$} & {$8.86\times 10$} \\ [0.5ex] 
 \hline\hline
  $\widehat{R^2}$ &  0.484&  0.483&  0.481&  0.477&  0.464\\
  Number in $\frak{I}_1$ & 14& 16& 13& 16& 16\\
Number in $\frak{I}_2$ &  160& 161& 161&160& 158 \\
 \hline
  \hline
$\lambda_l$ & {$1.44\times 10^{2}$} & {$2.34\times 10^{2}$} & {$3.79\times 10^{2}$} & {$6.16\times 10^{2}$} & {$1.00\times 10^{3}$} \\ [0.5ex] 
 \hline\hline
  $\widehat{R^2}$  &      0.442&  0.364&  0.133& -0.001& -0.001\\
Number of $\tilde{R}^2$ in $CI(\lambda_l)$ &17& 13&16&  8&  8\\
Number of $\tilde{R}^2$ in $CI_2(\lambda_l)$ &  155& 149& 149& 175& 175 \\
 \hline

\end{tabular}
  \caption{Number of $\tilde{R}^2$ in the confidence interval for Lasso.}
  \label{tab:R2_count_lasso}
\end{table}

\subsection*{Ridge.}

\begin{figure}[H]
  \begin{center}
    \includegraphics[width=0.6\columnwidth]{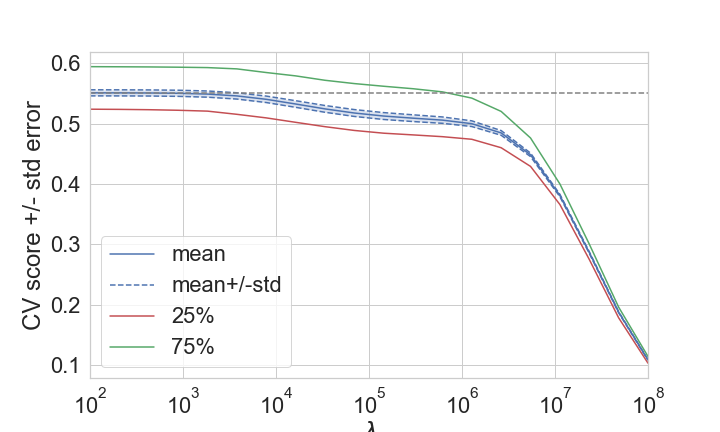}
    \caption{\label{fig:cross_validation_l2} Cross-validation score for Ridge.}
  \end{center}
\end{figure}

 \begin{table}[H]
  \centering\captionsetup{margin = 2.5cm}
  \setlength{\tabcolsep}{8pt}
\begin{tabular}{||c| c c c c c ||} 
 \hline
$\lambda_r$& {$1.00\times 10^{2}$  }     &     { $2.07\times 10^{2}$ }  &   {$4.28\times 10^{2}$ }&  { $8.85\times 10^{2}$   }  & { $1.83\times 10^{3}$}\\ [0.5ex] 
 \hline\hline
  $\widehat{R^2}$ & 0.551 & 0.551& 0.551& 0.550 & 0.549 \\
Number in $\frak{I}_1$ &  9& 9&  11&  9&  11\\
Number in $\frak{I}_2$ & 168& 167& 166& 166& 166\\
 \hline
 \hline
 $\lambda_r$& {$3.79\times 10^{3}$}  &  {$7.84\times 10^{3} $}   & {$1.62\times 10^{4}  $}  & {$3.36\times 10^{4}$}  & {$ 6.95\times 10^{4}$} \\ [0.5ex] 
 \hline\hline
  $\widehat{R^2}$ &   0.546& 0.540 & 0.533& 0.524& 0.518\\
Number in $\frak{I}_1$ & 11& 14& 15& 13& 11\\

Number in $\frak{I}_2$  &  168& 169& 170& 172& 173\\
 \hline
  \hline
$\lambda_r$ & {$1.44\times 10^{5}$}  &    {$2.98\times 10^{5}$} &   {$ 6.16\times 10^{5} $}&  {$1.27\times 10^{6}$}&  {$ 2.64\times 10^{6}$} \\ [0.5ex] 
 \hline\hline
  $\widehat{R^2}$ &0.513 & 0.509 & 0.506 & 0.500 & 0.485\\
Number in $\frak{I}_1$  & 11& 11& 13&17& 11\\

Number in $\frak{I}_2$  &  172& 171& 171&167& 164\\
 \hline
  \hline
$\lambda_r$ & {$5.46\times 10^{6}$} &  {$ 1.23\times 10^{7} $}& {$ 2.34\times 10^{7}$} &  {$4.83\times 10^{7} $}& {$ 1.00\times 10^{8}$}\\ [0.5ex] 
 \hline\hline
   $\widehat{R^2}$ &  0.448 & 0.381 & 0.287 & 0.188 & 0.109\\
Number in $\frak{I}_1$ & 8&  8& 6&  7& 7\\

Number in $\frak{I}_2$ &  159& 149& 143& 136& 136\\
 \hline
\end{tabular}
  \caption{Number of $\tilde{R}^2$ in the confidence interval for Ridge regression.}
  \label{tab:R2_count_ridge}
\end{table}

\subsection*{EN.}
\begin{figure}[H]
  \begin{subfigure}[b]{0.33\textwidth}
    \includegraphics[width=\textwidth]{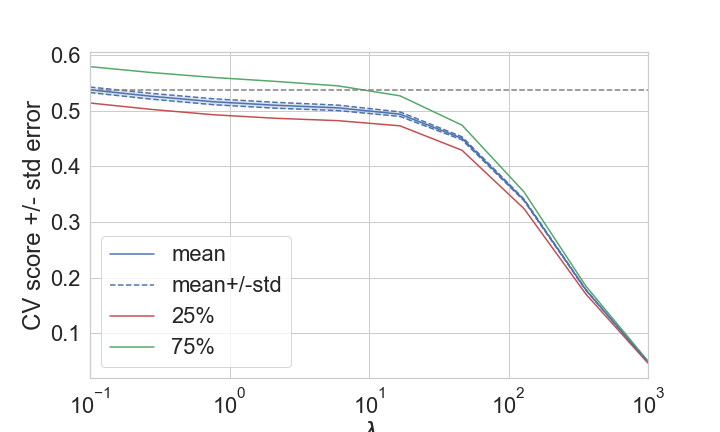}
    \caption{$\alpha=0.2$}
    \label{fig:cv_en_1}
  \end{subfigure}
  \begin{subfigure}[b]{0.33\textwidth}
    \includegraphics[width=\textwidth]{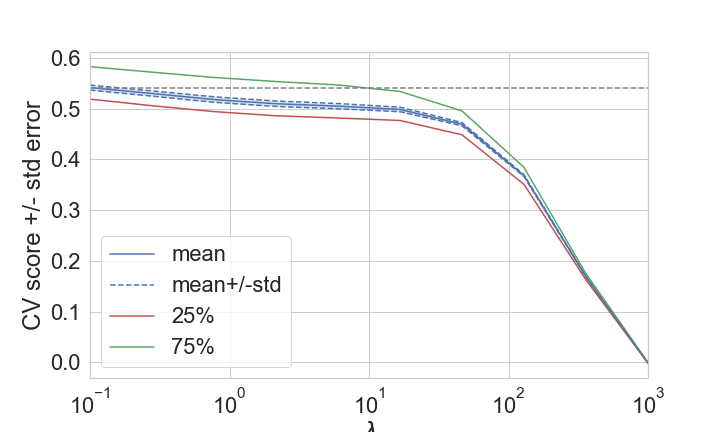}
    \caption{$\alpha=0.5$}
    \label{fig:cv_en_2}
  \end{subfigure}
  \begin{subfigure}[b]{0.33\textwidth}
    \includegraphics[width=\textwidth]{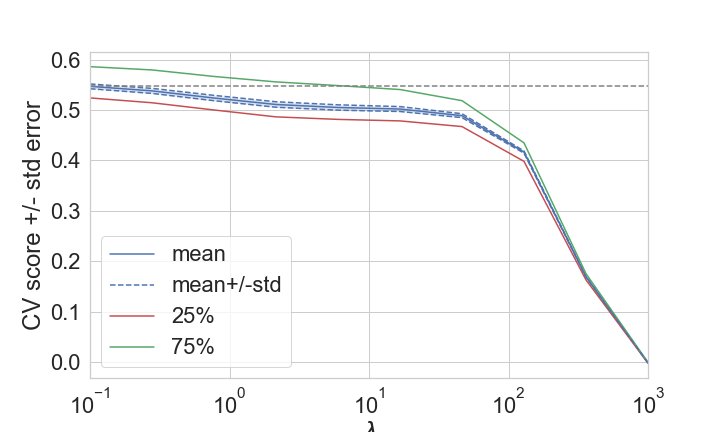}
    \caption{$\alpha=0.8$}
    \label{fig:cv_en_3}
  \end{subfigure}
  \caption{ Cross-validation score for EN.}\label{fig:cross_validation_EN}
\end{figure}

\begin{table}[H]
  \centering
  \setlength{\tabcolsep}{0.3pt}
\begin{tabular}{||c| c c c c c  ||} 
 \hline
$\alpha=0.2, \lambda_e=$ \quad\quad & \quad {$1.0\times 10^{-1}$} \quad&   \quad  {$2.78\times 10^{-1}$} \quad& \quad{$7.74\times 10^{-1}$} \quad&         \quad2.15 \quad&\quad5.99\quad \\ [0.5ex] 
 \hline
   $\widehat{R^2}$ &0.537 & 0.526& 0.516& 0.510& 0.505\\
$\frak{I}_1$ & 15& 16& 15& 13& 12\\

$\frak{I}_2$ &161& 160& 160& 160& 159 \\
\hline
$\alpha=0.2, \lambda_e=$ &       $ 1.68\times 10$&  { $4.64\times 10$} &       {$   1.29\times 10^{2}$}  & {$3.59\times 10^{2}$} & {$1.00\times 10^{3}$} \\ [0.5ex] 
 \hline
   $\widehat{R^2}$ &  0.494& 0.450& 0.340& 0.178& 0.048\\
$\frak{I}_1$ &  9&  2& 16& 11& 13\\
$\frak{I}_2$ & 152& 150& 133& 137& 142 \\
 \hline
  \hline
$\alpha=0.5,\lambda_e=$& {$1.0\times 10^{-1}$} &      { $ 2.78\times 10^{-1}$} &  {$ 7.74\times 10^{-1}$}&         2.15 &5.99  \\ [0.5ex] 
 \hline
$\widehat{R^2}$ &  0.542&  0.530&  0.518&  0.5010&  0.505\\
$\frak{I}_1$ &16& 16& 17& 11& 12\\

$\frak{I}_2$ &160& 162& 163& 161& 160 \\

  \hline
$\alpha=0.5,\lambda_e=$ &    {  $ 1.68\times 10$} &   {$4.64\times 10$ }&      {$    1.29\times 10^{2}$}  &{$3.59\times 10^{2}$}&{$1.00\times 10^{3} $}\\ [0.5ex] 
 \hline
$\widehat{R^2}$ &  
        0.498&  0.470&  0.368&  0.169& -0.001\\
$\frak{I}_1$ &  13&  9& 17& 12& 16\\

$\frak{I}_2$ & 158& 154& 154& 138& 174 \\
 
 \hline
  \hline
$\alpha=0.8,\lambda_e=$ & {$1.0\times 10^{-1}$} &     {$   2.78\times 10^{-1}$}& {$  7.74\times 10^{-1}$}&         2.15 &5.99\\ [0.5ex] 
 \hline
 $\widehat{R^2}$ & 0.547& 0.538&  0.524&  0.508&  0.504 \\
$\frak{I}_1$  & 16& 16& 15& 17& 12\\
 
$\frak{I}_2$ & 159& 160& 160& 162& 159 \\
 \hline
 $\alpha=0.8,\lambda_e=$&    {  $ 1.68\times 10$}& { $ 4.64\times 10$}&       {$   1.29\times 10^{2} $} & {$3.59\times 10^{2}$}&{$1.00\times 10^{3}$}   \\ [0.5ex] 
 \hline
 $\widehat{R^2}$ &  0.502&  0.487&  0.417&  0.170 & -0.001\\
$\frak{I}_1$  &  13& 10&  9& 12& 16\\
 
$\frak{I}_2$ &  160& 158& 147& 139& 174 \\
 \hline
\end{tabular}
  \caption{Number of $\tilde{R}^2$ in confidence interval for EN.}
  \label{tab:R2_count_enet}
\end{table}

\subsection{Comparison between the Enhanced TRACE and  Standard TRACE datasets}\label{sec:two_trace_comparison}
The comparison between the Enhanced TRACE  and  Standard TRACE datasets for OLS and two-step LASSO is summarized here. 

\begin{figure}[H]
  \begin{center}
    \includegraphics[width=0.6\columnwidth]{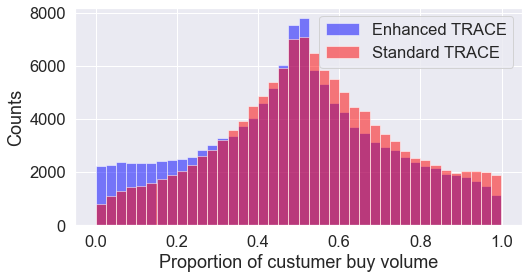} \centering
    \caption{\label{fig:propo_buy_enhanced_standard}  Proportion of customer buy orders in dollars (mean 0.48 for Enhanced TRACE and mean 0.52 ofr Standard TRACE).}
  \end{center}
\end{figure}

Table \ref{OLS_trace_comparison} compares the OLS regression with data in  Enhanced TRACE  and  Standard TRACE from the same period January 01, 2015-December 31, 2016. 
It shows similar $R^2$ values, with $55.4\%$ for Enhanced TRACE and $54.4\%$ for Standard TRACE. The relative difference between the regression coefficients is  small except for the following features.
\begin{itemize}
    \item {\it Prop Volume sell \$} and {\it Prop Volume buy \$}: For Enhanced TRACE, the coefficient of {\it Prop Volume buy \$} is 50\% larger than that of  {\it Prop Volume sell \$}. This relationship is reversed for Standard TRACE.  Further studies of the distribution of capped transactions show that 65\% of these transactions are customer sell orders. This contributes to  the changes of these two features in the Standard TRACE as well as  the difference in regression coefficients. See  Figure \ref{fig:propo_buy_enhanced_standard}.
    \item {\it Turnover} and {\it Years to maturity}: The coefficients of these two features are much bigger for Standard TRACE. This difference is tolerable as neither of two features is significant $(p\ge0.1)$.
\end{itemize}

Comparing the two-step LASSO model on these two datasets (see Table \ref{LASSO_standard_enhanced})  confirms similar conclusions: the outstanding features are consistent  and the regression coefficients are compatible.  

\begin{table}[H]
\centering
\small{
\centering\setlength{\tabcolsep}{4pt}
\captionsetup{margin=-2.cm}
\begin{tabular}{@{}l
*{2}{S[table-format=4.4,table-number-alignment=center]}@{}}\toprule
                    &   {Enhanced TRACE} & {Standard TRACE}   \\ \midrule
Volatility            &      80.7218{***}                   & 72.96{***}\\
                       &        {(0.253)      }               &{(0.21)} \\

 {Trading activity}            &         42.1450{***}        &    45.46{***}              \\
                     &            { (0.451)  }    &     { (0.496)  }     \\
 Log(Total Volume )              &        -21.1246{***}       &       -19.39{***}          \\
              &         {(0.252)  }         &      {   (0.261)  }           \\
      Years since issuance                               &       0.25{***}             &        0.32{***}       \\
           &             {(0.052)  }           &      {   (0.041) }       \\
Constant         &     86.1737          &     63.25      \\ \hline
N                                                                        &    152,408                      &130,716 \\
$R^2$                       &      52.8\%            &       52.5\%      \\ \bottomrule
\hline
\end{tabular}
\\ 
\scriptsize{Standard errors in parenthesis. Significance levels: * p$<$0.1, ** p$<$0.05, *** p$<$0.01. Two-tailed test.} \\
\scriptsize{Source: Enhanced TRACE (2015-2016).}
}
\caption{Two-step Lasso regression table: the impact  on bid-ask spread (in bp).\label{LASSO_standard_enhanced}}
\end{table}

\begin{table}[H]
\centering\captionsetup{margin = 2.5cm}
\small{
\centering\setlength{\tabcolsep}{4pt}
\begin{tabular}{@{}l
*{2}{S[table-format=4.4,table-number-alignment=center]}@{}}\\ \toprule 
                       {}              &  {TRACE Enhanced} &  {Standard TRACE} \\ \midrule
\hline
Volatility                              &   77.73{***}     &66.4{***} \\
 Number of trade days                      &         -3.66{***}   &  -2.85{***}     \\

Prop number of buys                          &         10.32{***}   &   9.93{***}      \\
					                  
      Prop number of sells                           &      32.45{***}  &    28.68{***}      \\
                                                               
 {Trading activity}                               &    46.32{***}   &  51.27{***}  \\
                                                                  
        Prop volume sell \$                            &   16.35{***}     & 23.18{***}    \\
                                                   
        Prop Volume buy \$                            &   26.42{***}   &   15.43{***}  \\
                                                                
 Log(total volume )                                   &    -21.40{***}    &  -20.62{***}      \\
                                                              
        Avg price                                          &   -0.12{***}  &    -0.14{***}       \\
                                                     
      Coupon                                             &   -0.47{***}   &   -0.42{***}          \\
                                                                 
      Duration                                           &         1.42{***}   &   1.98{***}   \\
                                                           
      Years to maturity                               &       -0.07   &    -0.37     \\
                                                        
      Years since issuance                                   &        1.26{***}    &  1.33{***}     \\
                                                       
      Turnover                                          &    -1.77   & -23.54                   \\

       LIBOR-OIS                                     &      34.01{***}    &  30.62{***}               \\
                                                             
Indicator of high yield bonds                  &   26.79{***}   &   23.49{***}         \\
                                                               
Indicator of investment grade bonds      &  15.99{***}   &   13.12{***}        \\
                                                              
Indicator of basic materials sector         &  8.90{***}   &  8.88{***}                \\
                                                             
Indicator of communications sector     &    5.13{***}   &  4.03{***}             \\
                                                             
Indicator of consumer, cyclical sector    & 4.03{***}   &   3.41{***}              \\
                                                               
Indicator of consumer, non-cyclical sector   &4.97{***}   &   4.40{***}     \\
                                                              
Indicator of energy sector                 &    3.88{***}   &   3.49{***}              \\
                                                              
Indicator of financial sector                    &   4.05{***}   &   2.99{***}                       \\
                                                              
Indicator of industrial sector                  &   2.815{***}    &  3.16{***}                   \\
                                                               
Indicator of technology sector            & 4.41{***}  &    4.06{***}        \\
                                                               
Indicator of utilities sector                &   4.58{***}   &   4.29{***}        \\

Constant                                                 &   42.77{***}    &  35.80{***}         \\ \hline
N                                                             &   152,408                     &                   130716         \\
$R^2$                                                     &    55.4\%                 &      54.4\%    \\
\bottomrule
\end{tabular}
\\ 
\scriptsize{Standard errors in parenthesis. Significance levels: * p$<$0.1, ** p$<$0.05, *** p$<$0.01. Two-tailed test. Source: TRACE Enhanced (2015-2016).} 
}
\caption{OLS regression: Comparison between TRACE Enhanced and Standard TRACE\label{OLS_trace_comparison}}
\end{table}

\section{Additional Details of the Price Impact Analysis}
\subsection{Estimation of TIM2 Model}\label{app:TIM2}
For simplicity, we omit the subscript $b$ f(or bond $b$) in the derivation here.
 Assume that there are two types of events $\Pi:=\{+1,-1\}$ with $+1$ denoting  customer-buy orders and $-1$ denoting  customer-sell orders. In this case, the mid-price dynamics \eqref{eq:tim_general} leads to the following expression for  mid-price changes:
\begin{eqnarray}
R_k(1):= M_{k+1}-M_t &=& \sum_{\pi \in \Pi}G_{\pi}(0)I(\pi_{k+1}=\pi)V_{k+1}^{\alpha}\epsilon_{k+1} + \eta_{k+1} \nonumber\\
&&+\sum_{j = 0}^{\infty}\sum_{\pi '\in \Pi} \underbrace{(G_{\pi'}(j +1) - G_{\pi'}(j))}_{\Delta_1 G_{\pi'}(j)}I(\pi_{k-j}=\pi') \epsilon_{k-j}V_{k-j}^{\alpha}.
\end{eqnarray}
As a consequence, we can write the conditional response functions and response correlation matrix as:
\begin{eqnarray}
S_{\pi}(l) &=& \mathbb{E}[R_{k}\cdot \epsilon_{k-l+1}|\pi_{k-l+1} = \pi]=\frac{\mathbb{E}[R_{k}\cdot \epsilon_{k-l+1}I(\pi_{k-l+1}=\pi)]}{\mathbb{P}(\pi)},\\
C_{\pi,\pi'} (n) &=& \mathbb{E}[\epsilon_t\epsilon_{t+n}V_{t+n}^{\alpha} | \pi_t=\pi, \pi_{t+n}=\pi^{\prime}]= \frac{\mathbb{E}[\epsilon_tI(\pi_t=\pi)\cdot V_{t+n}^{\alpha}\epsilon_{t+n}I(\pi_{t+n}=\pi')]}{\mathbb{P}(\pi_{t}=\pi,\pi_{t+n}=\pi')},
\end{eqnarray}
for $1 \leq l \leq L$ and $-N \leq n \leq L$. Then the response function can be written as, for $\pi=+1$ or $-1$:
\begin{eqnarray}
S_{\pi}(l) = \sum_{\pi '\in \Pi}\mathbb{P}(\pi'|\pi)G_{\pi'}(0)C_{\pi,\pi'}(l) +\sum_{j=0}^{+\infty}\sum_{\pi '\in \Pi}\mathbb{P}(\pi'|\pi)\Delta_1G_{\pi'}(j)\cdot C_{\pi,\pi'}(l-j-1).
\end{eqnarray}
Denote $\widetilde{C}_{\pi,\pi'}(l) =\mathbb{P}(\pi_{t+l} = \pi'|\pi_t=\pi) C_{\pi,\pi'}(l)$ and  $\widetilde{S}_{\pi}(l) =S_{\pi}(l) -\sum_{\pi '\in \Pi}G_{\pi'}(0)\widetilde{C}_{\pi,\pi'}(l) $ , then:

{\tiny\begin{eqnarray}
\underbrace{
\begin{pmatrix}
\widetilde{S}_{+1}(1)\\
\widetilde{S}_{+1}(2)\\
      \vdots\\
\widetilde{S}_{+1}(L)\\
\widetilde{S}_{-1}(1)\\
\widetilde{S}_{-1}(2)\\
      \vdots\\
\widetilde{S}_{-1}(L)
\end{pmatrix}}_{\overline{S}(L)}
\underbrace{
= {\begin{bmatrix} 
\widetilde{C}_{+1,+1}(0) & \widetilde{C}_{+1,+1}(-1) &\cdots & \widetilde{C}_{+1,+1}(-N+1)&\widetilde{C}_{+1,-1}(0) & \widetilde{C}_{+1,-1}(-1) &\cdots & \widetilde{C}_{+1,-1}(-N+1)\\
\widetilde{C}_{+1,+1}(1) & \widetilde{C}_{+1,+1}(0)  & \cdots & \widetilde{C}_{+1,+1}(-N+2)&\widetilde{C}_{+1,-1}(1) & \widetilde{C}_{+1,-1}(0) & \cdots & \widetilde{C}_{+1,-1}(-N+2)\\
\vdots&\vdots &\ddots& \vdots&\vdots&\vdots &\ddots& \vdots\\
\widetilde{C}_{+1,+1}(L-1) &\cdots&\cdots & \widetilde{C}_{+1,+1}(-N+L)&\widetilde{C}_{+1,-1}(L-1) &\cdots&\cdots & \widetilde{C}_{+1,-1}(-N+L)\\
\widetilde{C}_{-1,+1}(0) & \widetilde{C}_{-1,+1}(-1) &\cdots & \widetilde{C}_{-1,+1}(-N+1)&\widetilde{C}_{-1,-1}(0) & \widetilde{C}_{-1,-1}(1) &\cdots & \widetilde{C}_{-1,-1}(N-1)\\
\widetilde{C}_{-1,+1}(1) & \widetilde{C}_{-1,+1}(0)  & \cdots & \widetilde{C}_{-1,+1}(-N+2)&\widetilde{C}_{-1,-1}(1) & C_{-1,-1}(0) & \cdots & \widetilde{C}_{-1,-1}(-N+2)\\
\vdots&\vdots &\ddots& \vdots&\vdots&\vdots &\ddots& \vdots\\
\widetilde{C}_{-1,+1}(L-1) &\cdots&\cdots & \widetilde{C}_{-1,+1}(-N+L)&\widetilde{C}_{-1,-1}(L-1) &\cdots&\cdots & \widetilde{C}_{-1,-1}(-N+L)
\end{bmatrix}}}_{\overline{\mathcal{C}}(N,L)}
\underbrace{
\begin{pmatrix}
\Delta_1G_{+1}(0)\\
\Delta_1G_{+1}(1)\\
      \vdots\\
\Delta_1G_{+1}(N-1)\\
\Delta_1G_{-1}(0)\\
\Delta_1G_{-1}(1)\\
      \vdots\\
\Delta_1G_{-1}(N-1)\\
\end{pmatrix}}_{\overline{\mathcal{G}}(N)},
\end{eqnarray}}
where we have:
\begin{eqnarray*}
&&S_{1}(l) = \mathbb{E}[R_k|\epsilon_{k-l+1}  = 1],\quad S_{-1}(l) =  -\mathbb{E}[R_k|\epsilon_{k-l+1}  =- 1],\\
&&\widetilde{C}_{1,-1}(n)= -\frac{\mathbb{P}(\pi_t=1,\pi_{t+n}=-1)\mathbb{E}\left[V_{t+n}^{\alpha}|\pi_t=1,\pi_{t+n}=-1\right]}{\mathbb{P}(\pi_t=1)},\\
&&\widetilde{C}_{-1,1}(n)=-\frac{\mathbb{P}(\pi_t=-1,\pi_{t+n}=1)\mathbb{E}\left[V_{t+n}^{\alpha}|\pi_t=-1,\pi_{t+n}=1\right]}{\mathbb{P}(\pi_t=-1)},\\
&&\widetilde{C}_{-1,-1}(n)=\frac{\mathbb{P}(\pi_t=-1,\pi_{t+n}=-1)\mathbb{E}\left[V_{t+n}^{\alpha}|\pi_t=-1,\pi_{t+n}=-1\right]}{\mathbb{P}(\pi_t=-1)},\\
&&\widetilde{C}_{1,1}(n)=\frac{\mathbb{P}(\pi_t=1,\pi_{t+n}=1)\mathbb{E}\left[V_{t+n}^{\alpha}|\pi_t=1,\pi_{t+n}=1\right]}{\mathbb{P}(\pi_t=1)}.
\end{eqnarray*}
The signature plot for TIM2 model~\cite{eisler2012price} can be similarly defined as:
\begin{eqnarray}\label{eq:app_signature_plot2}
l \, D_{{\rm TIM2}}(l) &=& \sum_{0\leq n < l}\sum_{+1}G_{+1}(l-n)^2P(+1) + \sum_{n>0}\sum_{+1}[G_{+1}(l+n) - G_{+1}(n)]^2 P(+1) \nonumber\\
&+& 2 \sum_{0 \leq n <n' <l}\sum_{+1,-1}G_{+1}(l-n)G_{-1}(l-n')C_{+1,-1}(n'-n)\nonumber\\
&+& 2 \sum_{0 < n < n' <l}\sum_{+1,-1}[G_{+1}(l+n)-G_{+1}(n)][G_{-1}(l+n')-G_{-1}(n')]C_{+1,-1}(n-n') \nonumber\\
&+&2\sum_{0 \leq n <l}\sum_{n'>0}\sum_{+1,-1}G_{+1}(l-n) [G_{-1}(l+n')-G_{-1}(n')]C_{-1, +1}(n'+n).
\end{eqnarray}

\end{document}